%% file: PRLdraft_g-function.tex
\documentclass[%
reprint,
 superscriptaddress,
 amsmath,amssymb,
 aps,
 nofootinbib,
 prb,
 physics
]{revtex4-2}

\usepackage{multirow}
\usepackage{bigdelim}
\usepackage[dvipdfmx]{graphicx}
\usepackage{color}
\usepackage{dcolumn}
\usepackage{physics}
\usepackage{mathtools}
\usepackage{bm}
\usepackage{array}
\usepackage{comment}
\usepackage{cancel}
\usepackage{here}
\usepackage[usenames,dvipsnames]{xcolor}
\usepackage{mathrsfs}
\usepackage{amsmath,amsthm}
\usepackage{enumerate}
\usepackage{empheq} 
\theoremstyle{definition}

\usepackage{xcolor} 
\definecolor{RoyalBlue}{rgb}{0.255, 0.412, 0.882}
\definecolor{DeepSkyBlue}{rgb}{0.0, 0.749, 1.0}
\definecolor{Crimson}{rgb}{0.862, 0.078, 0.235}
\definecolor{ForestGreen}{rgb}{0.133, 0.545, 0.133}
\definecolor{OrangeRed}{rgb}{1.0, 0.271, 0.0}
\definecolor{Orchid}{rgb}{0.855, 0.439, 0.839}
\definecolor{Sienna}{rgb}{0.627, 0.322, 0.176}
\definecolor{Goldenrod}{rgb}{0.855, 0.647, 0.125}
\definecolor{CadetBlue}{rgb}{0.372, 0.619, 0.627}
\definecolor{CornflowerBlue}{rgb}{0.392, 0.584, 0.929}
\definecolor{RebeccaPurple}{rgb}{0.4, 0.2, 0.6}
\definecolor{Salmon}{rgb}{0.980, 0.502, 0.447}
\definecolor{HotPink}{rgb}{1.0, 0.412, 0.706}
\definecolor{Chocolate}{rgb}{0.824, 0.412, 0.118}
\definecolor{SteelBlue}{rgb}{0.275, 0.510, 0.706}
\definecolor{FireBrick}{rgb}{0.698, 0.133, 0.133}
\definecolor{bondiblue}{rgb}{0.0, 0.58, 0.71}
\definecolor{celestialblue}{rgb}{0.29, 0.59, 0.82}
\definecolor{coolblack}{rgb}{0.0, 0.18, 0.39}
\definecolor{frenchblue}{rgb}{0.0, 0.45, 0.73}
\definecolor{lapislazuli}{rgb}{0.15, 0.38, 0.61}
\definecolor{mediumpersianblue}{rgb}{0.0, 0.4, 0.65}
\definecolor{darkpowderblue}{rgb}{0.0, 0.2, 0.6}
\definecolor{darkcandyapplered}{rgb}{0.64, 0.0, 0.0}
\definecolor{darkscarlet}{rgb}{0.34, 0.01, 0.1}
\definecolor{falured}{rgb}{0.5, 0.09, 0.09}
\definecolor{darkcyan}{rgb}{0.0, 0.55, 0.55}

\newcommand{\ri}{\mathrm{i}}
\def\beq{\begin{equation}}
\def\eeq{\end{equation}}
\def\bp{\begin{pmatrix}}
\def\ep{\end{pmatrix}}

\usepackage[pdftex,colorlinks,urlcolor=darkcyan,citecolor=blue,linkcolor=blue]{hyperref}

\begin{document}

\author{Yi-Jun He}
\affiliation{School of physics \& Shing-Tung Yau Center, Southeast University, Nanjing  211189, P. R. China}

\author{Yunfeng Jiang}
\email{jinagyf2008@seu.edu.cn}
\affiliation{School of physics \& Shing-Tung Yau Center, Southeast University, Nanjing  211189, P. R. China}
\affiliation{Peng Huanwu Center for Fundamental Theory, Hefei, Anhui 230026, China}

\date{\today}

\title{Exact $g$-function without strings}
\begin{abstract}
We propose a new approach to compute exact $g$-function for integrable quantum field theories with non-diagonal scattering S-matrices. The approach is based on an integrable lattice regularization of the quantum field theory. The exact $g$-function is encoded in the overlap of the integrable boundary state and the ground state on the lattice, which can be computed exactly by Bethe ansatz. In the continuum limit, after subtracting the contribution proportional to the volume of the closed channel, we obtain the exact $g$-function, given in terms of the counting function which is the solution of a nonlinear integral equation. The resulting $g$-function contains two parts, the scalar part, which depends on the boundary parameters and the ratio of Fredholm determinants, which is universal. This approach bypasses the difficulties of dealing with magnetic excitations for non-diagonal scattering theories in the framework of thermodynamic Bethe ansatz. We obtain numerical and analytical results of the exact $g$-function for the prototypical sine-Gordon theory with various integrable boundary conditions.

\end{abstract}

\maketitle

\section{Introduction}
\label{sec:intro}
Boundary entropy, or exact $g$-function plays a fundamental role in systems with defects and boundaries. Originally it was introduced by Affleck and Ludwig in the study of Kondo problem \cite{Affleck:1991tk} to measure the boundary degrees of freedom of critical systems. In string theory, it describes the tension of a D-brane \cite{Harvey:1999gq}. The $g$-function is a boundary analog of Zamolodchikov's $c$-function, which decreases monotonically along boundary RG flow with fixed critical bulk theory. This is known as the $g$-theorem. It was conjectured in \cite{Affleck:1991tk} and proved in \cite{Friedan:2003yc}. In recent years, the $g$-theorem has been revisited from quantum information point of view \cite{Casini:2016fgb,Casini:2022bsu,Harper:2024aku}.

When the bulk theory is non-critical, the off-critical $g$-function is also a highly interesting quantity, especially for integrable quantum field theories (IQFTs). For many IQFTs in 1+1 dimensions with various integrable boundaries, the $g$-function can be computed exactly \cite{Woynarovich:2004gc,Dorey:2004xk,Dorey:2005ak,Dorey:2009vg,Dorey:2010ub,Pozsgay:2010tv,Caetano:2021dbh,Caetano:2020dyp}. More recently, the exact $g$-function makes its appearance in AdS/CFT correspondence. It was found that a class of OPE coefficients involving giant gravitons and a single-trace non-BPS operator in the planar $\mathcal{N}=4$ super-Yang-Mills theory is given by a worldsheet $g$-function \cite{Jiang:2019xdz,Jiang:2019zig}. This result should be generalizable to more classes of OPE coefficients in $\mathcal{N}=4$ SYM and ABJM theories which are given by the overlap of an energy eigenstate and an integrable boundary state \cite{Komatsu:2020sup,Kristjansen:2023ysz,Ivanovskiy:2024vel,Yang:2021hrl,Kristjansen:2021abc,Yang:2021kot,Jiang:2023cdm}. Such overlaps are essentially the $g$-functions and their excited state generalizations. The fact that such generalizations have not been achieved so far is largely due to our inadequate understanding of the computation of $g$-functions for \emph{non-diagonal scattering} theories.

The computation of exact $g$-functions for IQFTs turns out to be quite a non-trivial problem. Different approaches including a direct cluster expansion \cite{Dorey:2004xk,Dorey:2009vg,Dorey:2005ak}, thermodynamic Bethe ansatz (TBA) \cite{LeClair:1995uf,Woynarovich:2004gc,Pozsgay:2010tv,Kostov:2019sgu} and a more rigorous matrix-tree theorem approach \cite{Kostov:2018dmi,Vu:2019qxt} have been developed. So far, consensus between different methods have been achieved for theories with diagonal scattering S-matrices. For {non-diagonal} scattering theories such as the sine-Gordon theory and the planar $\mathcal{N}=4$ SYM, exact calculation of the $g$-function is still an outstanding open question. In \cite{Vu:2019qxt}, it was found that the TBA approach lead to divergences. The authors propose to regularize the TBA and compute the ratio of UV and IR $g$-functions, which is finite. This is not completely satisfactory as it still renders the individual $g$-function divergent. Therefore, it is crucial to develop alternative approaches to compute the $g$-function. This is the goal of the current work.

We focus on the prototypical example of non-diagonal scattering IQFT which is the sine-Gordon (sG) theory. We exploit an integrable lattice regularization of the sG theory and compute the partition function on a finite cylinder with integrable boundaries at the two ends. In the closed channel, the boundaries are represented by boundary states. From the definition of the $g$-function, it is naturally encoded in the overlap of the ground state and the boundary state. Such an overlap can be computed exactly on the lattice by Bethe ansatz. In the continuum limit, the overlap contains two parts, one is proportional to the circumference of the cylinder $R$, which is naturally identified with the contribution from the boundary energy; the other non-extensive part is precisely the $g$-function that we are after.

\section{Setups}
\label{sec:setup}

\subsection{Exact $g$-function}
Let us recall the definition of the $g$-function. Consider an IQFT on a finite cylinder with circumference $R$ and height $L$. The partition function can be computed in two different ways. In the open channel,
\begin{align}
Z_{ab}(R,L)=\tr e^{-RH_{ab}(L)}=\sum_n e^{-R E_{ab}^{(n)}(L)}
\end{align}
where $H_{ab}(L)$ is the Hamiltonian with two boundaries, which we assume to be integrable \cite{Ghoshal:1993tm}. 
In the closed channel, the integrable boundary conditions become boundary states and the partition function reads
\begin{align}
Z_{ab}(R,L)=&\,\langle B_a|e^{-H(R)L}|B_b\rangle\\\nonumber
=&\,\sum_n \langle B_a|n\rangle\langle n|B_b\rangle e^{-E^{(n)}(R)L}\,.
\end{align}
where $H(R)$ is the Hamiltonian with periodic boundary condition. In the $L\gg 1$ limit
\begin{align}
\label{eq:closedvac}
Z_{ab}(R,L)\approx\langle B_a|0\rangle\langle 0|B_b\rangle e^{-E^{(0)}(R)L}
\end{align}
where $E^{(0)}(R)\approx \mathcal{E}R+\mathcal{O}(1/R)$ is the ground state energy.
$\langle B_i|0\rangle$ contains a factor $e^{-\varepsilon_iR}$ $(i=a,b)$ where $\varepsilon_i$ is the boundary energy. The exact $g$-function is defined to be the non-extensive part of the overlap
\begin{align}
g_i(r)=e^{\varepsilon_iR}\langle B_i|0\rangle,\qquad i=a,b\,,
\end{align}
which only depends on $r=mR$ where $m$ is the mass scale.

\subsection{Sine-Gordon theory with two boundaries}
We consider the sG theory with two boundaries located at $x=x_{\pm}$. The action is given by
\begin{align}
\label{eq:action}
\mathcal{A}=\int_{-\infty}^{\infty}\mathrm{d}y\int_{x_-}^{x_+}\mathrm{d}x\mathcal{L}(\varphi)+\int_{-\infty}^{\infty}\mathcal{B}_\pm(\varphi)\mathrm{d}y\,
\end{align}
where the bulk and boundary Lagrangian densities are
\begin{align}
\label{eq:bulkL}
\mathcal{L}(\varphi)=\frac{1}{2}(\partial_{\mu}\varphi)^2+\mu_{\text{bulk}}\cos(\beta\varphi)
\end{align}
and
\begin{align}
\label{eq:boundaryL}
\mathcal{B}_{\pm}(\varphi)=\left.\mu_{\pm}\cos\left(\frac{\beta}{2}(\varphi-\varphi_0^{\pm})\right)\right|_{x=x_{\pm}}\,.
\end{align}
It has been shown by Ghoshal and Zamolodchikov that these boundary conditions are integrable \cite{Ghoshal:1993tm}. The quantum theory is described by the bulk and boundary S-matrices, which can be obtained non-perturbatively by the bootstrap approach \cite{Zamolodchikov:1978xm,Ghoshal:1993tm}. The explicit expressions of the S-matrices can be found in Sec.~\ref{app:bootstrap} of SM. The bulk and boundary S-matrices depend on a set of parameters $\lambda$ and $(\zeta_\pm,\vartheta_\pm)$ respectively. These IR parameters are related to parameters of the UV description, \emph{i.e.} the action \eqref{eq:action} by the UV-IR relation \cite{Bajnok:2001ug} 
\begin{align}
\label{eq:bulkUV-IR}
\lambda=\frac{8\pi}{\beta^2}-1=\frac{1}{\nu-1}  \,.
\end{align}
The UV-IR relation for boundary parameters is given in Sec.~\ref{app:bootstrap}. Here we have introduced another parameter $\nu$ in \eqref{eq:bulkUV-IR} for later convenience.

\subsection{Light cone six-vertex model}
\begin{figure}[h!]
\centering
\includegraphics[scale=0.21]{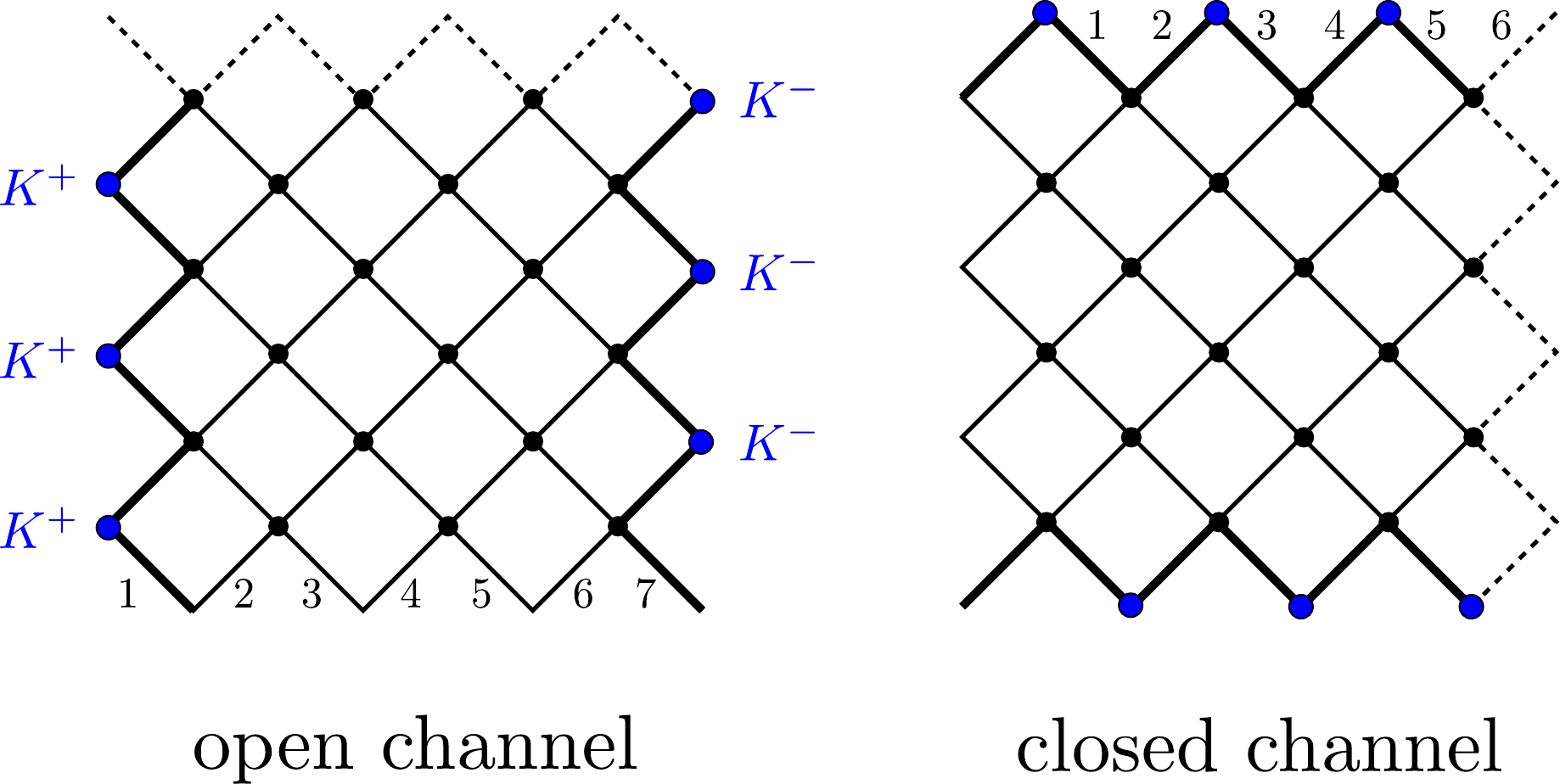}
\caption{The light cone six-vertex model with integrable boundaries. The left and right panel correspond to the open and closed channels respectively.}
\label{fig:DtoD}
\end{figure}

It is well-known that sG theory can be regularized on a light cone lattice as an integrable lattice model \cite{Destri:1992qk,Destri:1994bv}. We consider such a regularization with boundaries, as is shown in Fig~\ref{fig:DtoD}. We consider a lattice of size $(2M+1)\times (2N)$ where $2M+1$ and $2N$ are the sizes of the open and closed channels respectively. 
For simplicity, from now on we take the two boundaries to be the same, \emph{i.e.} $|B_a\rangle=|B_b\rangle$. At each bulk vertex, there are six non-zero configurations whose Boltzmann weights is given by the $\check{R}$-matrix
\begin{align*}
\check{R}_{ij}(u)=\left(\begin{array}{cccc} \sinh(u+\ri\gamma) & 0 & 0 &0\\ 0& \ri\sinh\gamma & \sinh u & 0\\ 0&\sinh u& \ri\sinh\gamma& 0\\ 0 & 0&0&\sinh(u+\ri\gamma)\end{array}\right)
\end{align*}
where $0<\gamma<\pi$ is the bulk parameter. In the open channel, vertices on the boundaries are described by the boundary $K$-matrices. They are given by $K^+(u)=K(u+\ri\gamma)$ and $K^-(u)=K(u)$ where $K(u)$ is a $2\times 2$ matrix with the following matrix elements
\begin{align*}
&K_{11}(u)=2\ri(\sin a\cosh b\cosh u+\cos a\sinh b\sinh u)\,,\\\nonumber
&K_{12}(u)=K_{21}(u)=\sinh 2u\,,\\\nonumber
&K_{22}(u)=2\ri(\sin a\cosh b\cosh u-\cos a\sinh b\sinh u)\,.
\end{align*}
The bulk $R$-matrix and boundary $K$-matrix satisfy the bulk and boundary Yang-Baxter equations respectively, which establishes integrability of the lattice model. 
The relation between the parameters of the lattice model $(\gamma,a,b)$ and those of the sine-Gordon model $(\nu,\zeta_{\pm}=\zeta,\vartheta_\pm=\vartheta)$  has been worked out in \cite{LeClair:1995uf,Ahn:2003ns} and is given by $\nu=\pi/\gamma$ and
\begin{align}
\label{eq:boundaryIRlattice}
\zeta=\frac{\pi}{2}\frac{\pi+2\ri a}{\pi-\gamma},\qquad\vartheta=\frac{\pi b}{\pi-\gamma}\,.
\end{align}


\section{Exact $g$-function}
\label{sec:BAE}

\subsection{Partition function in the closed channel}
We compute the lattice partition function in the closed channel, which is depicted in the right panel of Fig.~\ref{fig:DtoD}. Such lattice partition function can be computed by the method in \cite{Bajnok:2020xoz} (see Sec.~\ref{app:lattice} of SM for details), yielding
\begin{align}
\label{eq:ZMNrewrite}
    Z_{M,N}(u)=\frac{\langle\Phi_0^+|U^{\dagger}\mathbf{T}_R(u)^M|\Phi_0^-\rangle}{(\mathrm{i}\sin\gamma)^{2MN}} ,  
\end{align}
where  $U^{\dagger}$ is the shift operator which shifts one site towards left, $\mathbf{T}_R(u)$ is the double-row transfer matrix
\begin{align}
\mathbf{T}_R(u)=&\,\widehat{\tau}(\tfrac{\tilde{u}}{2};\{\pm\tfrac{\tilde{u}}{2}\})\tau(\tfrac{\tilde{u}}{2};\{\pm\tfrac{\tilde{u}}{2}\})
\end{align}
with
\begin{align*}
    \tau(\tfrac{\tilde{u}}{2};\{\pm\tfrac{\tilde{u}}{2}\})=&\,\text{tr}_a R_{a1}(0)R_{a2}(\tilde{u})\ldots R_{a,2N-1}(0)R_{a,2N-1}(\tilde{u}) \,,\\
    \widehat{\tau}(\tfrac{\tilde{u}}{2};\{\pm\tfrac{\tilde{u}}{2}\})=&\,\text{tr}_a R_{a,2N}(0)R_{a,2N-1}(\tilde{u})\ldots R_{a2}(0)R_{a1}(\tilde{u})  \,.
\end{align*}
Here $\tilde{u}=-u-\ri\gamma$ and $R_{ij}(u)=P_{ij}\check{R}_{ij}(u)$ with $P_{ij}$ being the permutation operator. The boundary states are given by
\begin{align}
|\Phi_0^-\rangle=&\,\bigotimes_{m=1}^N\left(\sigma^x\check{K}^-(\tfrac{u}{2})\right)_{ij}(-1)^{i-1}|i\rangle_{2m-1}\otimes|j\rangle_{2m}\,,\\\nonumber
\langle\Phi_0^+|=&\,\bigotimes_{m=1}^N\left(\check{K}^+(\tfrac{-u}{2})\sigma^x\right)_{ij}(-1)^{j-1}{_{2m-1}}\langle i|\otimes{_{2m}}\langle j|\,,
\end{align}
where $\check{K}^+_m(\tfrac{-u}{2})=\text{tr}_a\left[K_a^+(\tfrac{u}{2})\check{R}_{am}(u)\right]$ and $\check{K}^-_m(\tfrac{u}{2})=K^-_m(\tfrac{u}{2})$. The indices $i,j=1,2$.

\subsection{Integrability}
Integrability of the lattice model allows us to derive analytic results for the partition function \eqref{eq:ZMNrewrite}. 
\paragraph{Bulk integrability} Due to bulk integrability, we can diagonalize the transfer matrix $\mathbf{T}_{R}(u)$ by Bethe ansatz, \emph{i.e.} we can construct the $K$-magnon Bethe state $|\mathbf{u}_K\rangle$ such that $\mathbf{T}_R(u)|\mathbf{u}_K\rangle=\tau_R(u|\mathbf{u}_K)|\mathbf{u}_K\rangle$ with eigenvalue $\tau_R(u|\mathbf{u})$ given by \cite{Yung:1994td,Bajnok:2020xoz}
\begin{align}
\tau_R(u)=\left(\ri \sinh u\sin\gamma\right)^{2N}\frac{Q(-\tfrac{u}{2}-\ri\gamma)}{Q(-\tfrac{u}{2})}\frac{Q(\tfrac{u}{2}+\ri\gamma)}{Q(\tfrac{u}{2})}\,,
\end{align}
where we have introduced the trigonometric $Q$-polynomial
\begin{align}
Q(u)=\prod_{k=1}^K\sinh(u-u_k)\,.
\end{align}
The Bethe roots $\mathbf{u}_K$ are solutions of the Bethe ansatz equation (BAE). After introducing the following quantity
\begin{align*}
\mathfrak{a}(v)=\left(\frac{\sinh(v+\tfrac{u}{2})\sinh(v-\tfrac{u}{2}-\ri\gamma)}{\sinh(v-\tfrac{u}{2})\sinh(v+\tfrac{u}{2}+\ri\gamma)}\right)^N\frac{Q(v+\ri\gamma)}{Q(v-\ri\gamma)}
\end{align*}
the BAE reads
\begin{align}
\mathfrak{a}(u_k)=-1,\qquad k=1,\ldots,K\,.
\end{align}
\paragraph{Boundary integrability} The states $|\Phi_0^\pm\rangle$ are integrable boundary states \cite{Piroli:2017sei} in the spin chain language. Their overlap with the Bethe state $|\mathbf{u}_K\rangle$ is non-vanishing only if the Bethe roots $\mathbf{u}_K$ are paired, namely $\mathbf{u}_K=\{u_1,-u_1,\ldots,u_{K/2},-u_{K/2}\}$ assuming $K$ is even\footnote{In the odd $K$ case, one of the Bethe roots is zero and the rest are paired.}. For the paired states, the overlap takes the form \cite{Pozsgay:2018ybn}
\begin{align*}
W(u|\mathbf{u})=&\,\frac{\langle\Phi_0^+|U^{\dagger}|\mathbf{u}\rangle\langle\mathbf{u}|\Phi_0^-\rangle}{\langle\mathbf{u}|\mathbf{u}\rangle}\\\nonumber
=&\,\left[-\sinh(u+2\ri\gamma)\sinh(u) \right]^{2N}\\\nonumber
&\times\frac{Q(-\tfrac{u}{2}-\ri\gamma)}{Q(-\tfrac{u}{2})}\prod_{j=1}^K f(u_j)\times\frac{\det G^+}{\det G^-}
\end{align*}
where 
\begin{align}
\label{eq:deff}
f(u)=\frac{4\sinh^2(u+\ri a)\cosh^2(u+b)}{\sinh(2u+\ri\gamma)\sinh(2u)}
\end{align}
and $G^{\pm}$ are $\tfrac{K}{2}\times\tfrac{K}{2}$ Gaudin-like matrices
\begin{align}
G_{jk}^\pm=\delta_{jk}\frac{\mathfrak{a}'(u_j)}{\mathfrak{a}(u_j)}-\left[\varphi(u_j-u_k)\pm\varphi(u_j+u_k)\right]
\end{align}
with
\begin{align}
\label{eq:varphimain}
\varphi(u)=-\frac{\ri\sin(2\gamma)}{\sinh(u+\ri\gamma)\sinh(u-\ri\gamma)}\,.
\end{align}
To sum up, the partition function \eqref{eq:ZMNrewrite} can be written as
\begin{align}
\label{eq:partitionZMN}
Z_{M,N}(u)=\frac{1}{(\ri\sin\gamma)^{2MN}}\sum_{\text{sol}}\tau_R(u|\mathbf{u})^M W(u|\mathbf{u})\,.
\end{align}
where we sum over all physical solutions of the BAE.
\paragraph{AFV state} In order to make contact with sG theory in the continuum limit, we take the spectral parameter $u=-2\Theta-\ri\gamma$. We will consider the ground state of the sG theory, which corresponds to the anti-ferromagnetic vacuum (AFV) state of the lattice model. Following \cite{Destri:1994bv}, we define the counting function
\begin{align}
Z_N(u)=&\,-\ri\ln\mathfrak{a}(u)\\\nonumber
=&\, N\left[\phi_{1/2}(u-\Theta)+\phi_{1/2}(u+\Theta)\right]-\sum_{j=1}^N\phi_1(u-u_j)
\end{align}
where we have introduced the function
\begin{align}
\phi_{x}(u)=\ri\ln\left(\frac{\sinh(\ri\gamma x+u)}{\sinh(\ri\gamma x-u)}\right)\,.
\end{align}
The AFV state corresponds to the following solution of BAE
\begin{align}
Z_N(u_k)=(-N+2k-1)\pi,\qquad k=1,\ldots,N\,.
\end{align}
\paragraph{Non-linear integral equation (NLIE)} Using the contour deformation trick, we can show that the counting function satisfy an nonlinear integral equation.
To obtain results for the sG theory, we take the continuum limit. We will use $\Delta$ to denote lattice spacing. In the continuum limit, we take $\Delta\to0$ and $N,\Theta\to\infty$ in such a way that
\begin{align}
R=N\Delta,\qquad m=\frac{4}{\Delta}e^{-\frac{\Theta\pi}{\gamma}}
\end{align}
are finite. Here $R$ is the circumference of the cylinder and $m$ is the soliton mass of sG theory. In this limit, we obtain the NLIE for sG theory \cite{Destri:1994bv}
\begin{align}
\label{eq:NLIE}
Z(u)=&\,mR\sinh(\frac{\pi u}{\gamma})\\\nonumber
&\,+2\text{Im}\int_{-\infty}^{\infty}\mathrm{d} v G(u-v-\ri\xi)\ln\left(1+e^{\ri Z(v+\ri\xi)}\right)
\end{align}
where $\xi$ is a small real number which shifts the contour slightly away from the real axis and the kernel $G(u)$ is given by
\begin{align}
G(u)=\int_{-\infty}^{\infty}\frac{\mathrm{d}k}{2\pi}e^{\ri k u}\frac{\sinh((\tfrac{\pi}{2}-\gamma)k)}{2\sinh((\pi-\gamma)\tfrac{k}{2})\cosh(\tfrac{\gamma k}{2})}\,.
\end{align}

\subsection{Exact $g$-function}
In the limit $L\gg 1$, the partition function \eqref{eq:partitionZMN} is dominated by the AFV state
\footnote{Strictly speaking, the lattice partition function \eqref{eq:ZMNrewrite} we obtain is Minkowskian and we need to perform a Wick rotation before taking the large $L$ limit where AFV state dominates. }. Comparing \eqref{eq:partitionZMN} with \eqref{eq:closedvac}, we expect that $\tau_{R}(\Theta|\mathbf{u}_{\text{AFV}})^M$ gives the bulk energy while $W(u|\mathbf{u}_{\text{AFV}})$ corresponds to $|\langle B_a|0\rangle|^2$ in the continuum limit. This is indeed what happens and we find that 
\begin{align}
\label{eq:boundaryEe}
&\frac{\ri}{\Delta}\ln\tau_{R}(\Theta|\mathbf{u}_{\text{AFV}})\to - \mathcal{E}R+\text{finite terms}+\cdots\,,\\\nonumber
&\ln W(\Theta|\mathbf{u}_{\text{AFV}})\to -2\varepsilon_a R+\text{finite terms}+\cdots\,
\end{align}
where the `finite terms' refers to the terms that remain finite when $R\to\infty$ and the ellipsis denote the UV divergent terms in the continuum limit. The derivation of \eqref{eq:boundaryEe} can be found in Sec.~\ref{app:bulk} of SM. The finite term contained in $W(\Theta|\mathbf{u}_{\text{AFV}})$ is precisely the $g$-function that we are after. The structure of the result is similar to the one obtained by TBA, consisting of two parts $\ln g_a=\ln g_{\text{pref}}+\ln g_{\text{det}}$. The prefactor part depends on the boundary parameters and is given by
\begin{align}
\label{eq:gpref}
\ln g_{\text{pref}}=&\,-\text{Im}\int_{-\infty}^{\infty}\frac{\mathrm{d}v}{2\pi}\,\tilde{f}(-v-\ri\xi)\ln\left(1+e^{\ri Z(v+\ri\xi)}\right)\\\nonumber
&+\texttt{discrete terms}
\end{align}
`\texttt{discrete terms}' comes from the contributions of poles and zeros of the prefactor $f(u)$ defined in \eqref{eq:deff}, we give their explicit forms in Sec.~\ref{app:gfunc} of SM. $\tilde{f}(u)$ is given by 
\begin{align}
\tilde{f}(v)=f(v)-\int_{-\infty}^{\infty} f(x)G(-x+v)\mathrm{d}x\,.
\end{align}
The determinant part which does not depend on boundary parameters reads
\begin{align}
\label{eq:gdet}
\ln g_{\text{det}}=\frac{1}{2}\ln\frac{\det(1-\hat{H}^+)}{\det(1-\hat{H}^-)}\,,
\end{align}
where the kernels $\hat{H}^{\pm}$ are defined by
\begin{align}
\hat{H}^{\pm}[y](x)=\int_{\Gamma}\frac{\mathrm{d}u}{2\pi\ri}\frac{\varphi(x-u)\pm\varphi(x+u)}{2(1+e^{-\ri Z(u)})}y(u)\,.
\end{align} 
$\varphi(u)$ has been defined in \eqref{eq:varphimain} and the contour $\Gamma=\{\mathbb{R}+\ri\xi\}\cup\{\mathbb{R}-\ri\xi\}$, with $0<\xi<\tfrac{\gamma}{2}$. The expression of $g$-function in terms of the counting function \eqref{eq:gpref} and \eqref{eq:gdet} is the main result of this work.

\section{Results}
\label{sec:generalizations}
In this section, we present results of the $g$-function based on our method. We first solve the NLIE \eqref{eq:NLIE} to find the counting function $Z(u)$ and then plug in \eqref{eq:gpref} and \eqref{eq:gdet} to compute the $g$-function. The NLIE in general can be solved numerically by iteration and the Fredholm determinants are computed by discretization. More details can be found in Sec.~\ref{app:num} of SM. The procedure gives finite results. In some special limits, we can obtain analytic results which will also be presented below.

\paragraph{Free fermion point}
The free fermion point is particularly simple and can serve as a consistency check. At the free fermion point $\gamma=\tfrac{\pi}{2}$ with Dirichlet boundary condition, our result of $\ln g$ is the same as TBA up to a constant shift
\begin{align}
    \ln |g|_{\text{(NLIE)}} = \ln |g|_{\text{(TBA)}} - \tfrac{1}{2} \ln 2 . 
\end{align}
Two comments are in order. Firstly, in our approach a different normalization of the boundary state can lead to a constant shift of $\ln |g|$. Meanwhile, in \cite{Pozsgay:2010tv} it has been pointed out that the TBA approach misses the factor $- \tfrac{1}{2} \ln 2 $ while in our approach we recover it naturally. For more details and discussions, we refer to Sec.~\ref{app:FreeF} of SM.  

\paragraph{UV and IR limits}
In the UV and IR limit, the $g$-function approaches to a constant. In the IR limit $m R\gg 1$, we have\footnote{For the free boundary condition, this result should be modified. See Sec.~\ref{app:IRUV} of SM for details. } 
\begin{align}
    \ln |g| \sim \tfrac{1}{4}\ln\left(\tfrac{1}{2}-\tfrac{\gamma}{2\pi}\right)  \,.
\end{align}
In the UV limit $m R\ll 1$, we have $\ln |g| \sim 0$. These results have been checked numerically and the derivations can be found in Sec.~\ref{app:IRUV} of SM.

\begin{figure}[h!]
\centering
\includegraphics[scale=0.45]{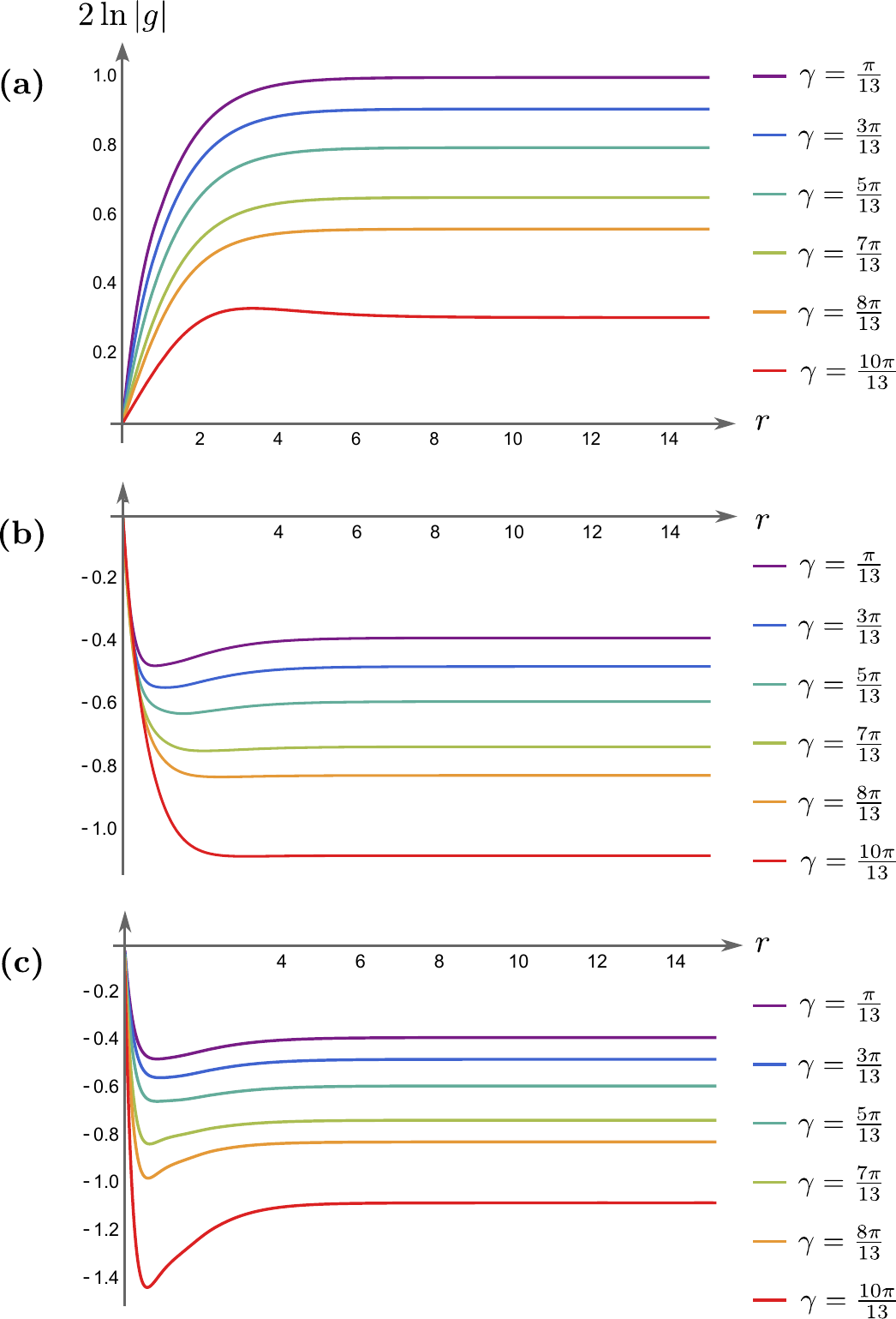}
\caption{$g$-function for different boundary conditions and $\gamma$ from UV to IR. We plot $2\ln|g|$ as a function of $r=Rm$ in the range $0\le r\le 15$. The boundary conditions in the three figures are chosen to be \textbf{(a)} Free boundary condition $a=b=0$; \textbf{(b)} A generic boundary condition with $a=b=1$; \textbf{(c)} Dirichlet boundary condition with $a=1$, $b=100$.}
\label{fig:UVtoIR}
\end{figure}

\paragraph{Bulk flow}
For generic values of bulk and boundary parameters, the $g$-function can be computed numerically. In figure~\ref{fig:UVtoIR}, we plot $\ln|g|^2$ as a function of $r=mR$ with different boundary and bulk parameters $a,b$ and $\gamma$. We see that $\ln|g|^2$ goes from 0 in the UV limit to a stable IR value. For different $\gamma$, the behaviors are qualitatively the same.
Notice that here we are changing the scale of the bulk, therefore the $g$-function does not have to change monotonically \cite{Green:2007wr}.

\begin{figure}[h!]
\centering
\includegraphics[scale=0.5]{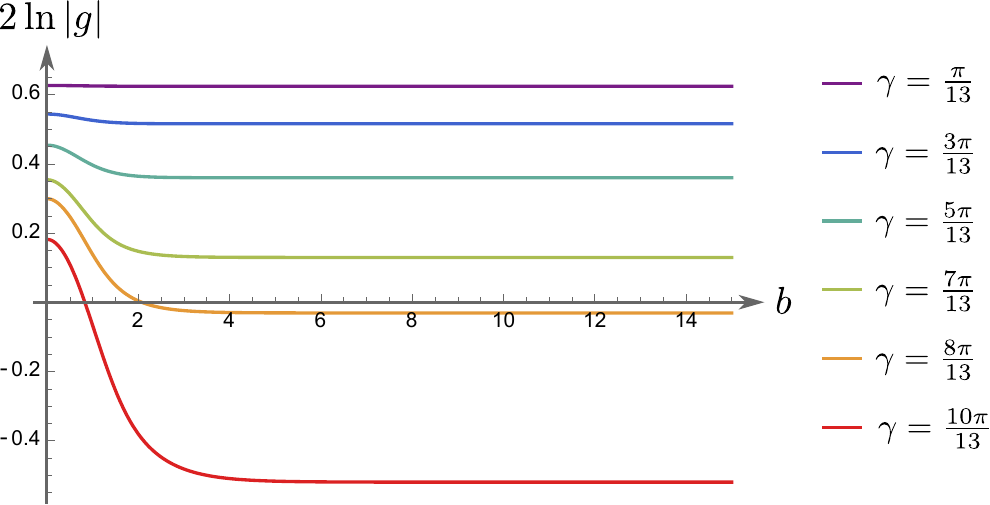}
\caption{Plot of $2\ln|g|$ as a function of $b$. We fix $a=0$, $r=1$ and take $b$ in the regime $0\le b\le 15$. The boundary condition flows from free boundary to fixed boundary conditions.}
\label{fig:F2D}
\end{figure}

\paragraph{Boundary flow} In Fig~\ref{fig:F2D}, we plot $\ln |g|^2$ as a function of $b$ with fixed $r$, which shows how the $g$-function changes from free to fixed boundary conditions. This corresponds to the boundary flow with fixed bulk theory and the $g$-function is decreasing monotonically, which is consistent with the $g$-theorem.

\section{Discussions}
In this work, we developed a new approach to compute the exact $g$-function of IQFT with non-diagonal scattering. Although our main focus is the sG model, the method is applicable to more general models which allows integrable lattice regularization, see \emph{e.g.} \cite{Zinn-Justin:1997iiv,Ikhlef:2008zz,Ikhlef:2011ay,Robertson:2020imc}.\par 

Our method has considerable conceptual and technical advantages compared to TBA. We work in the closed channel and extract the $g$-function from the overlap of ground state and the boundary state, which follow directly from the definition, without the need to {go} to the mirror channel and {make} string hypothesis.
For numerical computations,
we only need to solve one NLIE while the TBA equation typically involve several or even infinitely many $Y$-functions. This is a huge simplification, especially for the $g$-function, since all the $Y$-functions enter the computation of Fredholm determinants.
Using this approach, we obtained the exact $g$-function for the sG theory for the first time.

Based on the current work, there are a number of important future directions to explore. A detailed comparison of our method with TBA approach would be desirable. For bulk free energy, equivalence between NLIE and TBA have been established for a number of models including XXZ spin chain \cite{Takahashi2001}, 1d Fermi Hubbard model \cite{Cavaglia:2015nta} and the sG theory \cite{Balog:2003xd}. For the $g$-function it is expected to be more subtle since a naive computation using TBA lead to divergences. In this case, establishing the relation between the counting function and the $Y$-functions and then rewriting our results in terms of the $Y$-functions may shed light on a proper modification of the TBA approach for the $g$-function.

One important future goal is the computation of worldsheet $g$-function which is essential for computing numerous types of exact correlation functions in planar $\mathcal{N}=4$ SYM theory \cite{Jiang:2019zig,Jiang:2019xdz,Komatsu:2020sup,Kristjansen:2023ysz,Ivanovskiy:2024vel} and ABJM theory \cite{Yang:2021hrl,Kristjansen:2021abc,Yang:2021kot,Jiang:2023cdm}. At the moment, the worldsheet $g$-function was derived in the TBA framework. The worldsheet scattering theory is non-diagonal and hence the resulting $g$-function is expected to suffer from the aforementioned complications. Our result indicates that it is possible to compute the $g$-function using NLIE-like methods, written in terms of finitely many unknown functions. 

To achieve this goal, it is necessary to generalize our method to models with higher rank global symmetries. One possible hint is noticing that
the computation of exact $g$-function in IQFT is technically quite similar to the computation of surface free energy in integrable lattice models \cite{Pozsgay:2018ybn,Kozlowski:2012fv,Gombor:2023bez,Gombor:2024api}. Recently, there has been considerable progress in the exact overlap formula of eigenstates and integrable boundary states for spin chains with higher rank symmetry \cite{deLeeuw:2016umh,Gombor:2022deb,Gombor:2022deb,Gombor:2021hmj,Gombor:2020kgu,Rylands:2022gev}, it is timely to study surface free energy for more general lattice models with higher rank global symmetries such as the $\mathfrak{gl}(m|n)$ invariant spin chains and the Hubbard model. This problem is not only interesting on its own right, but will also provide us with useful lessons on how to formulate the $g$-function in the framework of quantum spectral curve \cite{Gromov:2013pga,Ekhammar:2021pys}.







\section*{Acknowledgment}
The work of Y.J. is partly supported by Startup Funding no. 3207022217A1 of Southeast University and by the NSF of China through Grant No. 12247103.
Y.H.'s work is supported by the Jiangsu Funding Program for Excellent Postdoctoral Talent. We are indebted to Zoltan Bajnok and Andrea Kl\"umper for very helpful discussions and correspondences. We also thank Ivan Kostov, Didina Serban and Shota Komatsu for helpful comments on the draft.

\bibliography{refs.bib}


\include{Supplement}

\end{document}

%% file: Supplement.tex
\onecolumngrid

\begin{appendix}

\begin{center}
	\textbf{{\large Supplemental Material}}
\end{center}

\section{Bootstrap description of boundary sine-Gordon theory}
\label{app:bootstrap}
In this appendix, we collect useful formulae of the bulk and boundary S-matrices of the sine-Gordon theory \cite{Ghoshal:1993tm}. The particle spectrum of the sine-Gordon theory depends on the value of parameter $\lambda$. In the repulsive regime $\lambda<1$, the spectrum only contains solitons and anti-solitons. In the attractive regime $\lambda>1$, there are $\lfloor \lambda \rfloor$ species of breathers if $\lambda$ is not integer, and $\lambda-1$ species of breathers if $\lambda$ is an integer. 

\subsection{Bulk S-matrix}
The scattering between (anti-)solitons is described by the following S-matrix \cite{Zamolodchikov:1978xm}:
\begin{align}
    S_{++}^{++}(\theta)=S_{--}^{--}(\theta)=S_0(\theta), \quad S_{+-}^{+-}(\theta)=S_{-+}^{-+}(\theta)=S_T(\theta) S_0(\theta) , \quad S_{+-}^{-+}(\theta)=S_{-+}^{+-}(\theta)=S_R(\theta) S_0(\theta) ,  \notag\\
    S_T(\theta)=\frac{\sinh \left(\lambda\theta\right)}{\sinh \left(\lambda(\ri \pi-\theta)\right)}, \quad S_R(\theta)=\frac{\ri \sin \left(\lambda\pi\right)}{\sinh \left(\lambda(\ri \pi-\theta)\right)}, \quad S_0(\theta)= - \exp \left( \int_{-\infty}^{\infty} \frac{\mathrm{d} t}{t} \frac{\sinh \left(\frac{t \pi}{2}(\lambda^{-1}-1)\right)}{2 \sinh \left(\frac{\pi  t}{2 \lambda }\right) \cosh \left(\frac{\pi t}{2}\right)} e^{\ri \theta t}\right),   
\end{align}
where the parameter $\lambda$ is related to the parameter $\beta$ in the UV description \eqref{eq:bulkL} by
\begin{align}
    \lambda=\frac{8\pi}{\beta^2}-1.  
\end{align}
The notation `$\pm$' stands for soliton/anti-soliton respectively. 

For convenience, we introduce the elementary block 
\begin{align}
    S_a(\theta)=[a]_\theta=\frac{\sinh \theta+\ri \sin \pi a}{\sinh \theta-\ri \sin \pi a} \,.  
\end{align}
Then the S-matrix between (anti-)solitons and breathers can be expressed as 
\begin{align}
    S_{ \pm, B_k}(\theta)=\prod_{a \in P_k}[a]_\theta \quad P_k=\left\{\frac{1-k \lambda^{-1}}{2}, \frac{1-(k-2) \lambda^{-1}}{2}, \ldots, \frac{1+(k-2) \lambda^{-1}}{2}\right\}
\end{align}
and the S-matrix between breathers can be expressed as
\begin{align}
    & S_{B_k, B_{k^{\prime}}}(\theta)=\prod_{a \in P_{k k^{\prime}}}[a]_\theta    \notag\\
    & P_{k k^{\prime}}=\left\{\frac{\left(k+k^{\prime}\right) \lambda^{-1}}{2}, \frac{\left(k+k^{\prime}-2\right) \lambda^{-1}}{2}, \frac{\left(k+k^{\prime}-2\right) \lambda^{-1}}{2}, \ldots, \frac{\left(\left|k-k^{\prime}\right|+2\right) \lambda^{-1}}{2}, \frac{\left(\left|k-k^{\prime}\right|+2\right) \lambda^{-1}}{2}, \frac{\left|k-k^{\prime}\right| \lambda^{-1}}{2}\right\} .
\end{align}

\subsection{Boundary S-matrix}
The (anti-)soliton boundary scattering amplitudes can be expressed by the matrix \cite{Ghoshal:1993tm,Bajnok:2001ug}
\begin{align}
    R_s(u) & =\left(\begin{array}{cc}
    P_0^{+}(\zeta, \vartheta, u) & Q_0(u) \\
    Q_0(u) & P_0^{-}(\zeta, \vartheta, u)
    \end{array}\right) R_0(u) \frac{\sigma(\zeta, u)}{\cos (\zeta)} \frac{\sigma(\ri \vartheta, u)}{\cosh (\vartheta)} \,, \\
    P_0^{ \pm}(\zeta, \vartheta, u) & =\cos (\lambda u) \cos (\zeta) \cosh (\vartheta) \mp \sin (\lambda u) \sin (\zeta) \sinh (\vartheta) \,,\\
    Q_0(u) & =-\sin (\lambda u) \cos (\lambda u) \,,
\end{align}
where $u=-\ri \theta$ and $\theta$ is the rapidity,  
\begin{align}
    R_0(u)=\prod_{l=1}^{\infty}\left[\frac{\Gamma\left(4 l \lambda-\frac{2 \lambda u}{\pi}\right) \Gamma\left(4 \lambda(l-1)+1-\frac{2 \lambda u}{\pi}\right)}{\Gamma\left((4 l-3) \lambda-\frac{2 \lambda u}{\pi}\right) \Gamma\left((4 l-1) \lambda+1-\frac{2 \lambda u}{\pi}\right)} /(u \rightarrow-u)\right]
\end{align}
is the boundary condition independent part and 
\begin{align}
    \sigma(x, u)=\frac{\cos x}{\cos (x+\lambda u)} \prod_{l=1}^{\infty}\left[\frac{\Gamma\left(\frac{1}{2}+\frac{x}{\pi}+(2 l-1) \lambda-\frac{\lambda u}{\pi}\right) \Gamma\left(\frac{1}{2}-\frac{x}{\pi}+(2 l-1) \lambda-\frac{\lambda u}{\pi}\right)}{\Gamma\left(\frac{1}{2}-\frac{x}{\pi}+(2 l-2) \lambda-\frac{\lambda u}{\pi}\right) \Gamma\left(\frac{1}{2}+\frac{x}{\pi}+2 l \lambda-\frac{\lambda u}{\pi}\right)} /(u \rightarrow-u)\right]
\end{align}
depends on the boundary condition. The singularities of this matrix corresponds to boundary bound states and boundary resonance states \cite{Skorik:1995ky,Mattsson:2000aw}. 

The boundary scattering amplitudes for breathers $B_n$ are \cite{Ghoshal:1993iq,Bajnok:2001ug}
\begin{align}
    \label{chap1.15}
    R^{(n)}(\zeta, \vartheta, u)=R_0^{(n)}(u) S^{(n)}(\zeta, u) S^{(n)}(\ri \vartheta, u) \,.
\end{align}
Using the following shorthand notation 
\begin{align}
    (x)=\frac{\sin \left(\frac{u}{2}+\frac{x \pi}{2}\right)}{\sin \left(\frac{u}{2}-\frac{x \pi}{2}\right)} \,,
\end{align}
we have
\begin{align}
    R_0^{(n)}(u)=\frac{\left(\frac{1}{2}\right)\left(\frac{n}{2 \lambda}+1\right)}{\left(\frac{n}{2 \lambda}+\frac{3}{2}\right)} \prod_{l=1}^{n-1} \frac{\left(\frac{l}{2 \lambda}\right)\left(\frac{l}{2 \lambda}+1\right)}{\left(\frac{l}{2 \lambda}+\frac{3}{2}\right)^2} \quad , \quad S^{(n)}(x, u)=\prod_{l=0}^{n-1} \frac{\left(\frac{x}{\lambda \pi}-\frac{1}{2}+\frac{n-2 l-1}{2 \lambda}\right)}{\left(\frac{x}{\lambda \pi}+\frac{1}{2}+\frac{n-2 l-1}{2 \lambda}\right)} \,.
\end{align}
The boundary S-matrices which describe the boundary scattering at $x=x_{\pm}$ depend on parameters $\zeta_\pm$, $\vartheta_\pm$. They are related to parameters of the UV description \eqref{eq:boundaryL} in the following way \cite{Bajnok:2001ug}
\begin{align}
\begin{split}
    &\cos\left(\frac{\beta^2\zeta_\pm}{8\pi}\right)\cosh\left(\frac{\beta^2\vartheta_\pm}{8\pi}\right)=\frac{\mu_\pm}{\mu_{\text{crit}}}\cos\left(\frac{\beta\phi^\pm_0}{2} \right)\,,\\
    &\sin\left(\frac{\beta^2\zeta_\pm}{8\pi}\right)\sinh\left(\frac{\beta^2\vartheta_\pm}{8\pi}\right)=\frac{\mu_\pm}{\mu_{\text{crit}}}\sin\left(\frac{\beta\phi^\pm_0}{2} \right)
\end{split}
\end{align}
where $\mu_{\text{crit}}=\sqrt{2\mu_{\text{bulk}}/\sin(\beta^2/8\pi)}$.
The bulk and boundary parameters $\lambda$ and $\zeta,\vartheta$ are related to the lattice parameters $\gamma$ and $a,b$ by relations\footnote{$\nu$ is related to $\lambda$ by the equation \eqref{app:IRUV}} $\nu=\pi/\gamma$ and \eqref{eq:boundaryIRlattice} in the main text.

\section{Details on lattice computations}
\label{app:lattice}
In this appendix, we present details on the computation of lattice partition function $Z_{M,N}(u)$.

\subsection{The Closed Channel Partition Function}
The lattice partition function of the light cone six-vertex model have been studied in \cite{Yung:1994td,Bajnok:2020xoz} for diagonal $K$-matrices. It is straightforward to generalize the computation to non-diagonal $K$-matrices. Following these works, the closed channel partition function is given by 
\begin{align}
    Z_{M,N}(u)=\langle\Psi_0^+\left(\tfrac{u}{2}\right)|U^{\dagger}(\mathbf{T}_D(u))^M |\Psi_0^- \left(\tfrac{u}{2}\right)\rangle
\end{align}
where $|\Psi_0^{\pm}(\tfrac{u}{2})\rangle$ are the two-site states 
\begin{align}
|\Psi_0^-(u)\rangle=&\,\otimes_{m=1}^N\left(\sigma^x\check{K}^-(u)\right)_{ij}\,|i\rangle_{2m-1}\otimes|j\rangle_{2m}\,,\\\nonumber
\langle\Psi_0^+(u)|=&\,\otimes_{m=1}^N\left(\check{K}^+(-u)\sigma^x\right)_{ij}\,{_{2m-1}}\langle i|\otimes{_{2m}}\langle j|\,.
\end{align}
with $\check{K}^+_m(\tfrac{-u}{2})=\text{tr}_a\left[K_a^+(\tfrac{u}{2})\check{R}_{am}(u)\right]$ and $\check{K}^-_m(\tfrac{u}{2})=K^-_m(\tfrac{u}{2})$. The two site states $|\Psi_0^\pm(u)\rangle$ come from rotating the boundary $K$-matrices in the open channel by 90$^\circ$.
The operator $\mathbf{T}_D(u)$ is the diagonal-to-diagonal transfer matrix in the closed channel 
\begin{align*}
    \mathbf{T}_D(u)=\tilde{V}^{(2)}(u)\tilde{V}^{(1)}(u)  
\end{align*}
where
\begin{align*}
    &\tilde{V}^{(1)}(u)=(-1)^N \Omega^{(2)} \check{R}_{23}(\tilde{u})\check{R}_{45}(\tilde{u})\ldots \check{R}_{2N,1}(\tilde{u}) \Omega^{(1)},  \\
    &\tilde{V}^{(2)}(u)=(-1)^N\Omega^{(1)} \check{R}_{12}(\tilde{u})\check{R}_{34}(\tilde{u})\ldots \check{R}_{2N-1,2N}(\tilde{u}) \Omega^{(2)}  
\end{align*}
with $\tilde{u}=-u-\ri\gamma$ being the `cross transformed' spectral parameter. $U$ is the one site shift operator defined by
\begin{align}
U=P_{12}P_{23}\cdots P_{2N-1,2N}\,.
\end{align}
The operators $\Omega^{(1,2)}$ are
\begin{align*}
    \Omega^{(1)}=\sigma_1^z\sigma_3^z\ldots\sigma_{2N-1}^z,\qquad \Omega^{(2)}=\sigma_2^z\sigma_4^z\ldots\sigma_{2N}^z . 
\end{align*}

To diagonalize the diagonal-to-diagonal transfer matrix $\mathbf{T}_D(u)$, the key idea is to relate it to the row-to-row transfer matrix with special choices of inhomogeneities. In the closed channel, it is defined by
\begin{align}
    \mathbf{T}_{R}(u)=&\,\widehat{\tau}(\tfrac{\tilde{u}}{2};\{\pm\tfrac{\tilde{u}}{2}\})\tau(\tfrac{\tilde{u}}{2};\{\pm\tfrac{\tilde{u}}{2}\}) \,,
\end{align}
where
\begin{align}
    \tau(u;\{\mathbf{\theta}\})&=\text{tr}_a R_{a1}(u-\theta_1)\ldots R_{a,2N}(u-\theta_{2N}) \,,\\
    \widehat{\tau}(u;\{\mathbf{\theta}\})&=\text{tr}_a R_{a,2N}(u+\theta_{2N})\ldots R_{a1}(u+\theta_1)
\end{align}
are the inhomogeneous row-to-row transfer matrices. We make the following choice of the inhomogeneities
\begin{align}
\{\theta\}=\{\pm \tfrac{\tilde{u}}{2}\}\equiv\{\tfrac{\tilde{u}}{2},-\tfrac{\tilde{u}}{2},\dots,\tfrac{\tilde{u}}{2},-\tfrac{\tilde{u}}{2}\} \,.
\end{align}

The diagonal-to-diagonal transfer matrix is then related to the row-to-row transfer matrix by
\begin{align}
    \mathbf{T}_D(u) = \frac{\Omega^{(1)}\mathbf{T}_{R}(u)\Omega^{(1)}}{(\mathrm{i}\sin\gamma)^{2N}} ,  
\end{align}
thus the closed channel partition function can be rewritten as 
\begin{align}
    Z_{M,N}(u)=\frac{\langle\Phi_0^+|U^{\dagger}\mathbf{T}_{R}(u)^M|\Phi_0^-\rangle}{(\mathrm{i}\sin\gamma)^{2MN}},   
\end{align}
where 
\begin{align*}
    |\Phi_0^-\rangle=\Omega^{(1)}|\Psi_0^-(\tfrac{u}{2})\rangle \,, \qquad
    \langle\Phi_0^+|=\langle\Psi_0^+(\tfrac{u}{2})|\Omega^{(2)} \,.   
\end{align*}
This is \eqref{eq:ZMNrewrite} in the main text.

\subsection{Eigenvalue of the transfer matrix}
The transfer matrix $\mathbf{T}_{R}(u)$ can be diagonalized by algebraic Bethe ansatz. Using the crossing symmetry of $R$-matrix, it can be shown that 
\begin{align}
\label{eq:crossing}
    \tau(\tilde{u};\{\theta\})=\widehat{\tau}(u;\{\theta\}) . 
\end{align}
We define the monodromy matrix $T_a^{(2N)}(u;\{\theta_j\})$ in the usual way
\begin{align*}
    T_a^{(2N)}(u;\{\theta\})=R_{a1}(u-\theta_1)R_{a2}(u-\theta_2)\cdots R_{a,2N}(u-\theta_{2N})=\left(\begin{array}{ll}
    A(u) & B(u) \\
    C(u) & D(u)
    \end{array}\right)_a .  
\end{align*}
The eigenstates of the transfer matrix $\tau(u;\{\theta_j\})=\tr_a T_a^{(2N)}(u;\{\theta\})$ is constructed by 
\begin{align*}
|\mathbf{u}_K\rangle=B(u_1-\tfrac{\mathrm{i}\gamma}{2})\ldots B(u_K-\tfrac{\mathrm{i}\gamma}{2})|\!\uparrow\rangle^{2N} 
\end{align*}
where the Bethe roots $\{u_j\}$ satisfy the Bethe ansatz equation 
\begin{align}
    \prod_{j=1}^{2N}\frac{\sinh(u_k-\theta_j+\tfrac{\mathrm{i}\gamma}{2})}{\sinh(u_k-\theta_j-\tfrac{\mathrm{i}\gamma}{2})}
    =\prod_{k\ne j}^K\frac{\sinh(u_k-u_j+\mathrm{i}\gamma)}{\sinh(u_k-u_j-\mathrm{i}\gamma)},\qquad k=1,2,\ldots,K.
\end{align} 
The corresponding eigenvalue of $\tau(u;\{\theta_j\})$ is given by 
\begin{align*}
    \Lambda_c(u;\{\theta\})=&\,\prod_{j=1}^{2N}
    \sinh(u-\theta_j+\mathrm{i}\gamma)
    \prod_{k=1}^K\frac{\sinh(u-u_k-\tfrac{\mathrm{i}\gamma}{2})}{\sinh(u-u_k+\tfrac{\mathrm{i}\gamma}{2})}    \nonumber\\
    &\,+\prod_{j=1}^{2N}\sinh(u-\theta_j)\prod_{k=1}^{K}\frac{\sinh(u-u_k+\tfrac{3\mathrm{i}\gamma}{2})}{\sinh(u-u_k+\tfrac{\mathrm{i}\gamma}{2})}  
\end{align*}
Using the relation \eqref{eq:crossing}, we can derive the eigenvalue of $\widehat{\tau}(u;\{\theta_j\})$ as
\begin{align*}
    \widehat{\Lambda}_c(u;\{\theta\})=&\,\prod_{j=1}^{2N}\sinh(u+\theta_j+\mathrm{i}\gamma)\prod_{k=1}^{K}\frac{\sinh(u+u_k-\tfrac{\mathrm{i}\gamma}{2})}{\sinh(u+u_k+\tfrac{\mathrm{i}\gamma}{2})}    \nonumber\\
    &\, +\prod_{j=1}^{2N}
    \sinh(u+\theta_j)
    \prod_{k=1}^K\frac{\sinh(u+u_k+\tfrac{3\mathrm{i}\gamma}{2})}{\sinh(u+u_k+\tfrac{\mathrm{i}\gamma}{2})}    . 
\end{align*}

To obtain the sine-Gordon theory in the continuum limit, we take $u=-2\Theta-\ri\gamma$. Then (by a slight abuse of notation)
\begin{align*}
    \widehat{\tau}(\tfrac{\tilde{u}}{2};\{\pm\tfrac{\tilde{u}}{2}\})\tau(\tfrac{\tilde{u}}{2};\{\pm\tfrac{\tilde{u}}{2}\}) = \widehat{\tau}(\Theta;\{\pm\Theta\})\tau(\Theta;\{\pm\Theta\}) 
\end{align*}
and the corresponding eigenvalues are
\begin{align*}
    \Lambda_c\big(\Theta;\{\pm\Theta\} \big)=&\,\left[ \sinh(2\Theta+\mathrm{i}\gamma)\sinh(\mathrm{i}\gamma)\right]^N
    \prod_{k=1}^K\frac{\sinh(u_k-\Theta+\tfrac{\mathrm{i}\gamma}{2})}{\sinh(u_k-\Theta-\tfrac{\mathrm{i}\gamma}{2})}\,,\\
    \widehat{\Lambda}_c\big(\Theta;\{\pm\Theta\}\big)
    =&\,\left[\sinh(2\Theta+\mathrm{i}\gamma)\sinh(\mathrm{i}\gamma)\right]^N
    \prod_{k=1}^K\frac{\sinh(u_k+\Theta-\tfrac{\mathrm{i}\gamma}{2})}{\sinh(u_k+\Theta+\tfrac{\mathrm{i}\gamma}{2})} .  
\end{align*}

\subsection{Exact overlap formula}

The states $|\Phi^\pm\rangle$ are integrable boundary states. Their overlap with the Bethe state $|\mathbf{u}_K\rangle$ is non-vanishing only if the Bethe roots $\mathbf{u}_K$ are paired. For even $K$, the overlap formula can be extracted from \cite{Pozsgay:2018ybn}. After taking $u=-2\Theta-\ri\gamma$, the result is 
\begin{align}
\label{eq:6vtxOverlapR}    \frac{\langle\textbf{u}|\Phi_{0}^{-}(-\Theta-\tfrac{\ri\gamma}{2})\rangle}{\sqrt{\langle\textbf{u}|\textbf{u}\rangle}}&=\sinh^{N}(2\Theta+\ri\gamma) \sqrt{\prod_{j=1}^{K/2}F(u^{+}_j)}\sqrt{\frac{\det G_{jk}^{+}}{\det G_{jk}^{-}}}, \\
\label{eq:6vtxOverlapL}
    \frac{\langle\Phi_{0}^{+}(-\Theta-\tfrac{\ri\gamma}{2})|U^{\dagger}|\textbf{u}\rangle}{\sqrt{\langle\textbf{u}|\textbf{u}\rangle}}&=\sinh^{N}(2\Theta+\ri\gamma)\sinh^{N}(2\Theta-i\gamma) \\\notag
    &\times\prod_{k=1}^{K}\frac{\sinh(u_k-\Theta+\tfrac{\ri\gamma}{2})}{\sinh(u_k-\Theta-\tfrac{\ri\gamma}{2})}\sqrt{\prod_{j=1}^{K/2}F(u^{+}_j)}\sqrt{\frac{\det G_{jk}^{+}}{\det G_{jk}^{-}}}    
\end{align}
where
\begin{align}
    F(u_j)=\frac{16\sinh^2(u_j+\ri a)\sinh^2(u_j-\ri a)\cosh^2(u_j+b)\cosh^2(u_j-b)}{\sinh(2u_j+\ri\gamma)\sinh(2u_j-\ri\gamma)\sinh^2(2u_j)}, 
\end{align}
and 
\begin{align}
    G_{j k}^{\pm}&=\delta_{j k} m_{j}-\left[\varphi\left(u_{j}^{+}-u_{k}^{+}\right) \pm \varphi\left(u_{j}^{+}+u_{k}^{+}\right)\right], \\
    m_{j}&=\frac{\mathfrak{a}^{\prime}\left(u_{j}\right)}{\mathfrak{a}\left(u_{j}\right)}, \\
    \varphi(u)&=-\frac{\ri\sin(2\gamma)}{\sinh(u+\ri\gamma)\sinh(u-\ri\gamma)}. 
\end{align}
The quantity $\mathfrak{a}(u)$ is defined as
\begin{align}
    \mathfrak{a}(u)=\left(\frac{\sinh(u-\Theta-\tfrac{\ri\gamma}{2})\sinh(u+\Theta-\tfrac{\ri\gamma}{2})}{\sinh(u+\Theta+\tfrac{\ri\gamma}{2})\sinh(u-\Theta+\tfrac{\ri\gamma}{2})}\right)^{N}\prod_{j=1}^{K}\frac{\sinh(u-u_j+\ri\gamma)}{\sinh(u-u_j-\ri\gamma)}  .   
\end{align}

\section{Derivation of $g$-function}
\label{app:gfunc}
In this appendix, we give details for the derivation of the non-linear integral equation (NLIE) and the exact $g$-function. 
\subsection{Non-linear Integral Equation}
For later convenience, we introduce the function 
\begin{align*}
    \phi_{\nu}(u):=\ri\ln\frac{\sinh(\ri\gamma\nu+u)}{\sinh(\ri\gamma\nu-u)},  
\end{align*}
which is analytic in the strip $|\operatorname{Im}u|\leq \operatorname{min}(\nu\gamma,\pi-\nu\gamma)$ for real $\gamma$. We define the counting function 
\begin{align}
    \label{eq:logCount}
    Z_N(u)\equiv N\left(\phi_{1/2}(u-\Theta)+\phi_{1/2}(u+\Theta)\right)-\sum_{j=1}^{N}\phi_{1}(u-u_j),
\end{align}
where $\{u_j\}$ are the Bethe roots. For anti-ferromagnetic vacuum (AFV) state, the BAE can be written in terms of $Z_N(u)$ as
\begin{align}
    Z_{N}(u_k)=(-N+2k-1)\pi, \qquad k=1,2,\dots, N. 
\end{align}
To derive the NLIE, we transform the summation over the Bethe roots on the right hand side of \eqref{eq:logCount} into a contour integral, following  \cite{Destri:1994bv}. For the AFV state, we can choose the contour $\Gamma$ such that it tightly encloses all the Bethe roots. This contour can be deformed into $\Gamma=\{\mathbb{R}+\ri\xi\}\cup\{\mathbb{R}-\ri\xi\}$ with $0<\xi<\gamma/2$. 

We take the derivative with respect to $u$ on both sides of \eqref{eq:logCount} for later convenience, which leads to
\begin{align}
    \label{eq:nlie1}
    Z_{N}'(u)=Nz_{0}'(u)-\oint\frac{dv}{2\pi \ri}\phi_{1}'(u-v)\frac{\mathrm{d}}{\mathrm{d}v}\ln(1+e^{\ri Z_{N}(v)}),
\end{align}
where 
\begin{align}
    z_{0}(u)=\phi_{1/2}(u-\Theta)+\phi_{1/2}(u+\Theta). 
\end{align}
For our choice of contour $\Gamma$, \eqref{eq:nlie1} can be written as
\begin{align}
    \label{eq:nlie2}
    Z_{N}'(u)+\int_{-\infty}^{+\infty}\frac{\mathrm{d}v}{2\pi} \phi_{1}'(u-v) Z_{N}'(v)=Nz_{0}'(u)&-\ri\int_{\mathbb{R}+\ri\xi}\frac{\mathrm{d}v}{2\pi}\phi_{1}'(u-v)\frac{\mathrm{d}}{\mathrm{d}v}\ln(1+e^{\ri Z_{N}(v)})\notag\\
    &+\ri\int_{\mathbb{R}-\ri\xi}\frac{\mathrm{d}v}{2\pi}\phi_{1}'(u-v)\frac{\mathrm{d}}{\mathrm{d}v}\ln(1+e^{-\ri Z_{N}(v)}). 
\end{align}
Acting $(\delta+\phi_1'/2\pi)^{-1}$ on both sides of \eqref{eq:nlie2}, we obtain
\begin{align}
    Z_{N}'(u)=&N\left(\frac{\pi}{\gamma}\left(\operatorname{sech}\left(\tfrac{\pi}{\gamma}(u+\Theta)\right)+\operatorname{sech}\left(\tfrac{\pi}{\gamma}(u-\Theta)\right)\right)\right) \notag\\
    &-\ri\int_{\mathbb{R}+i\xi}\mathrm{d}v G(u-v)\frac{\mathrm{d}}{\mathrm{d}v}\ln(1+e^{\ri Z_{N}(v)})  +  \ri\int_{\mathbb{R}-\ri\xi}\mathrm{d}v G(u-v)\frac{\mathrm{d}}{\mathrm{d}v}\ln(1+e^{-\ri Z_{N}(v)}), 
\end{align}
where
\begin{align}
    G(u)=\int_{-\infty}^{+\infty}\frac{\mathrm{d}k}{2\pi}e^{\ri ku}\frac{\sinh((\tfrac{\pi}{2}-\gamma)k)}{2\sinh((\pi-\gamma)\tfrac{k}{2})\cosh(\tfrac{\gamma k}{2})}. 
\end{align}
Integrating over the variable $u$, we obtain
\begin{align}
    \label{eq:nile3}
    Z_{N}(u)=&N\left[\operatorname{gd}\left(\tfrac{\pi}{\gamma}(u+\Theta)\right)+\operatorname{gd}\left(\tfrac{\pi}{\gamma}(u-\Theta)\right)\right] \notag\\
    &-\ri\int_{\mathbb{R}+\ri\xi}\mathrm{d}v  G(u-v)\ln(1+e^{\ri Z_{N}(v)})   +  \ri\int_{\mathbb{R}-\ri\xi}\mathrm{d}v  G(u-v)\ln(1+e^{-\ri Z_{N}(v)}), 
\end{align}
where $\operatorname{gd}(x)\equiv\arctan(\sinh(x))$ is called hyperbolic amplitude (or Gudermannian). 
Taking the continuum limit $\Delta\to0$ and $N,\Theta\to\infty$ with
\begin{align}
\label{eq:contlimit}
R=N\Delta,\qquad m=\frac{4}{\Delta}e^{-\frac{\Theta\pi}{\gamma}}
\end{align}
fixed, \eqref{eq:nile3} becomes
\begin{align}
    Z(u)=mR\sinh\left(\frac{\pi u}{\gamma}\right)&-\ri\int_{\mathbb{R}+\ri\xi}\mathrm{d}v  G(u-v)\ln(1+e^{\ri Z(v)}) +  \ri\int_{\mathbb{R}-\ri\xi}\mathrm{d}v  G(u-v)\ln(1+e^{-\ri Z(v)}) , 
\end{align}
which is \eqref{eq:NLIE} in the main text.


\subsection{The $g$-Function}
\label{app:deriveG}
On the lattice, the squared $g$-function $|g|^2$ comes from the following overlap: 
\begin{align}
    \frac{\langle\Phi_{0}^{+}(-\Theta-\tfrac{\ri\gamma}{2})|U^{\dagger}|\textbf{u}\rangle\langle\textbf{u}|\Phi_{0}^{-}(-\Theta-\tfrac{\ri\gamma}{2})\rangle}{\langle\textbf{u}|\textbf{u}\rangle}=&\sinh^{N}(2\Theta-\ri\gamma)\sinh^{2N}(2\Theta+\ri\gamma)\\\nonumber
    &\,\times\prod_{k=1}^{N}\frac{\sinh(u_k-\Theta+\tfrac{\ri\gamma}{2})}{\sinh(u_k-\Theta-\tfrac{\ri\gamma}{2})}\prod_{j=1}^{N/2}F(u^{+}_j)\frac{\det G_{jk}^{+}}{\det G_{jk}^{-}}. 
\end{align}
On the right hand side, the factor $\sinh^{N}(2\Theta-\ri\gamma)\sinh^{2N}(2\Theta+\ri\gamma)$ is an $\{u_j\}$ independent normalization factor of the boundary states. The factor $\prod_{k=1}^{N}\frac{\sinh(u_k-\Theta+i\gamma/2)}{\sinh(u_k-\Theta-i\gamma/2)}$ is a pure phase which does not affect the $g$-function in the field theory limit. Therefore we focus on the continuum limit of
\begin{align}
\label{eq:scalarFactor}    \prod_{j=1}^{N/2}F(u^{+}_j)\frac{\det G_{jk}^{+}}{\det G_{jk}^{-}},    
\end{align}
Let us introduce the function $f(u)$ by the following relation
\begin{align}
    F(u)=f(u)f(-u) , 
\end{align}
where 
\begin{align}
    f(u)=\frac{4\sinh^2(u+\ri a)\cosh^2(u+b)}{\sinh(2u+\ri\gamma)\sinh(2u)}.  
\end{align}
The squared $g$-function $|g|^2$ is given by the field theory limit of this quantity after subtracting the contribution of the boundary energy and the UV divergence. 

For AFV state, the product $\prod_{j=1}^{N}f(u_j)$ can be rewritten into the contour integral, with the contour $\Gamma=\{\mathbb{R}+\ri\xi\}\cup\{\mathbb{R}-\ri\xi\}$. We take $a$ and $b$ to be real, and choose $\xi$ to be sufficiently small such that $\xi<a<\pi$, $0<\xi<\tfrac{\pi}{4}$ and $2\xi<\gamma<\pi$ are satisfied, then we have  
\begin{align}
\label{app:gfunc1}
    \frac{d}{du}\ln\prod_{a=1}^{N}f(u-u_a)=&\int_{\Gamma}\frac{dv}{2\pi \ri}\frac{d}{du}\left(\ln f(u-v)\right)\frac{d}{dv}\ln(1+e^{\ri Z_N(v)}) +\frac{d}{du}\ln(1+e^{\ri Z_N(u)}).
\end{align}
The derivative here is taken for later convenience. The first term on the right hand side can be rewritten as 
\begin{align}
    &\int_{\Gamma}\frac{dv}{2\pi \ri}\frac{f'(u-v)}{f(u-v)}\frac{d}{dv}\ln(1+e^{\ri Z_N(v)})  \notag\\
    =&\int_{\mathbb{R}-\ri\xi}\frac{dv}{2\pi}\frac{f'(u-v)}{f(u-v)}Z_N'(v)  \notag\\
    &+\ri\int_{\mathbb{R}+\ri\xi}\frac{dv}{2\pi} \frac{d}{du}\left(\frac{f'(u-v)}{f(u-v)}\right)\ln(1+e^{\ri Z_N(v)}) -\ri\int_{\mathbb{R}-\ri\xi}\frac{dv}{2\pi}\frac{d}{du}\left(\frac{f'(u-v)}{f(u-v)}\right)\ln(1+e^{-\ri Z_N(v)})  .  
\end{align}
After integrating over $u$, the first term on the right hand side of this equation becomes that 
\begin{align}
\label{eq:firstC7}
    \int_{\mathbb{R}-\ri\xi}\frac{dv}{2\pi}\ln f(u-v)Z_N'(v). 
\end{align}
Using the NLIE \eqref{eq:nile3},
\eqref{eq:firstC7} becomes 
\begin{align}
\label{eq:tmpC18}
    &\int_{\mathbb{R}-\ri\xi}\frac{dv}{2\pi}\ln f(u-v)Z_N'(v)  \notag\\
    =& \frac{N}{2\gamma} \int_{-\infty}^{+\infty} dx \ln f(u-x+\ri\xi)  \left(\operatorname{sech}\left(\tfrac{\pi}{\gamma}(x+\Theta-\ri\xi)\right)+\operatorname{sech}\left(\tfrac{\pi}{\gamma}(x-\Theta-\ri\xi)\right)\right)  \notag\\
    & -\ri\int_{-\infty}^{+\infty}\frac{dv\, dx}{2\pi} \frac{d}{du}(\ln f(u-x+\ri\xi)) G(x-v-2 \ri\xi) \ln(1+e^{\ri Z_N(v+\ri\xi)}) \notag\\
    & +\ri \int_{-\infty}^{+\infty}\frac{dv\, dx}{2\pi}\frac{d}{du}(\ln f(u-x+\ri\xi)) G(x-v) \ln(1+e^{-\ri Z_N(v-i\xi)}). 
\end{align}
Note that we have performed the integration by parts in the above equation. We can move the integration contour of $x$ of the second term on the right hand side of \eqref{eq:tmpC18} by $2\ri\xi$, then the pole of $(\ln f)'(u)$ at $u=0$ will contribute an extra term, leading to
\begin{align}
    &\int_{-\infty}^{+\infty}\frac{dx}{2\pi} \frac{d}{du}\ln f(u-x+\ri\xi) G(x-v-2 \ri\xi)  
    =\int_{-\infty}^{+\infty}\frac{dx}{2\pi} \frac{d}{du}(\ln f(u-x-\ri\xi)) G(x-v)  -\ri G(u-v-\ri\xi)  .   
\end{align}
Therefore after integration, the equation \eqref{app:gfunc1} becomes that
\begin{align}
\label{app:prefact0}
    \ln\prod_{a=1}^{N}f(u-u_a) 
    =& \frac{N}{2\gamma} \int_{-\infty}^{+\infty} dx \ln f(u-x+\ri\xi)  \left(\operatorname{sech}\left(\tfrac{\pi}{\gamma}(x+\Theta-\ri\xi)\right)+\operatorname{sech}\left(\tfrac{\pi}{\gamma}(x-\Theta-\ri\xi)\right)\right)  \notag\\
    &-\frac{1}{\pi} \operatorname{Im}\left[ \int_{-\infty}^{+\infty} dv \, dx (\ln f)'(-x-\ri\xi)  \,  (\delta-G)(x-v) \ln(1+e^{\ri Z_N(v+\ri\xi)}) \right] \notag\\
    &+\int_{-\infty}^{+\infty}dv G(u-v-\ri\xi)\ln(1+e^{\ri Z_N(v+\ri\xi)}) -\ln(1+e^{\ri Z_N(u)})  . 
\end{align}
{As explained in appendix \ref{app:BoundaryEnergy2}, the first term on the right hand side is proportional to $R$, \emph{i.e.} it belongs to the extensive part and does not contribute to the $g$-function}. For the rest terms, the field theory limit can be taken by replacing the function $Z_N(u)$ by $Z(u)$. Thus in the field theory limit, the term $\ln\prod_{j=1}^{N}f(u_j)$ will contribute the following terms to the logarithm of $g$-function: 
\begin{align}
   \ln g_{\text{pref}}= &-\frac{1}{2\pi} \operatorname{Im} \int_{-\infty}^{+\infty} \frac{dv \, dx}{2\pi} (\ln f)'(-x-\ri\xi)  \,  (\delta-G)(x-v) \ln(1+e^{\ri Z(v+\ri\xi)})  \notag\\
    &+\frac{1}{2}\int_{-\infty}^{+\infty}dv G(-v-\ri\xi)\ln(1+e^{\ri Z(v+\ri\xi)})  -\frac{1}{2}\ln(1+e^{\ri Z(0)}) . 
\end{align}
The term $\tfrac{1}{2}\ln\left(\frac{\det G_{jk}^{+}}{\det G_{jk}^{-}}\right)$ will contribute the following term to the logarithm of $g$-function \cite{Pozsgay:2018ybn}: 
\begin{align}
\label{eq:DetExpan}
 \ln g_{\text{det}}=  &\frac{1}{2}\ln\left(\frac{\operatorname{det}(1-\hat{H}^{+})}{\operatorname{det}(1-\hat{H}^{-})}\right)  \notag\\
    =&-\sum_{n=1}^{\infty} \frac{1}{2n}\left(\prod_{j=1}^n \oint_{\mathcal{C}} \frac{\mathrm{d} z_j}{2 \pi \ri} \frac{\mathfrak{a}\left(z_j\right)}{1+\mathfrak{a}\left(z_j\right)}\right) \varphi\left(z_1+z_2\right) \varphi\left(z_2-z_3\right) \ldots \varphi\left(z_n-z_1\right) .
\end{align}
where $\hat{H}^{ \pm}$ act as
\begin{align}
\label{eq:Hoperat}
    \left(\hat{H}^{ \pm}(h)\right)(x)=\oint_{\mathcal{C}} \frac{\mathrm{d} u}{2 \pi \ri} \frac{\mathfrak{a}(u)}{\mathfrak{a}(u)+1} \frac{\varphi(x-u) \pm \varphi(x+u)}{2} h(u)
\end{align}
for an arbitrary function $h(u)$, and the function $\varphi(u)$ is given by
\begin{align}
\label{eq:varphi}
    \varphi(u)=-\frac{\ri\sin (2 \gamma)}{\sinh (u+\ri\gamma) \sinh (u-\ri\gamma)}  .  
\end{align}
Thus the full result of $g$-function in the field theory limit is 
\begin{align}
\label{eq:appC2gfunc}
\begin{split}
  \ln g =  &- \operatorname{Im} \int_{-\infty}^{+\infty} \frac{dv \, dx}{2\pi} (\ln f)'(-x-\ri\xi)  \,  (\delta-G)(x-v) \ln(1+e^{\ri Z(v+\ri\xi)})    +   \frac{1}{2}\ln\left(\frac{\operatorname{det}(1-\hat{H}^{+})}{\operatorname{det}(1-\hat{H}^{-})}\right)   \\
    &+\frac{1}{2}\int_{-\infty}^{+\infty}dv G(-v-i\xi)\ln(1+e^{\ri Z(v+\ri\xi)})  -\frac{1}{2}\ln(1+e^{\ri Z(0)}) .  
\end{split}
\end{align}
Note that this result is valid for $\xi<a<\pi$, $0<\xi<\tfrac{\pi}{4}$ and $2\xi<\gamma<\pi$. The terms in the second line are the `\texttt{discrete terms}' in \eqref{eq:gpref} of the main text. For generic choices of parameters $\gamma$, $a$ and $\xi$, the `\texttt{discrete terms}' can be different. The general expression is 
\begin{align}
\begin{split}
  \ln g =  &- \operatorname{Im} \int_{-\infty}^{+\infty} \frac{dv \, dx}{2\pi} (\ln f)'(-x-\ri\xi)  \,  (\delta-G)(x-v) \ln(1+e^{\ri Z(v+\ri\xi)})    +   \frac{1}{2}\ln\left(\frac{\operatorname{det}(1-\hat{H}^{+})}{\operatorname{det}(1-\hat{H}^{-})}\right)   \\
    &-\frac{1}{2}\sum_{a=1}^{N_Z}\int_{-\infty}^{+\infty}dv G(-z_a-v-i\xi)\ln(1+e^{iZ(v+i\xi)})  +   \frac{1}{2}\sum_{a=1}^{N_P}\int_{-\infty}^{+\infty}dv G(-w_a-v-i\xi)\ln(1+e^{iZ(v+i\xi)}) \\
    &+\frac{1}{2}\sum_{a=1}^{N_Z}\ln(1+e^{iZ(-z_a)})-\frac{1}{2}\sum_{a=1}^{N_P}\ln(1+e^{iZ(-w_a)}) .  
\end{split}
\end{align}
The sets $\{w_a;a=1,\dots,N_P\}$ and $\{z_a;a=1,\dots,N_Z\}$ denote poles and zeros of the function $f(u)$ encircled in the contour $\Gamma$. In particular, for the free boundary condition where $a=b=0$, the expression for the $g$-function should be 
\begin{align}
\begin{split}
    \ln g_{\text{free}}= &- \operatorname{Im} \int_{-\infty}^{+\infty} \frac{dv \, dx}{2\pi}  (\ln f)'(-x-\ri\xi)  \,  (\delta-G)(x-v) \ln(1+e^{\ri Z(v+\ri\xi)})   +  \frac{1}{2}\ln\left(\frac{\operatorname{det}(1-\hat{H}^{+})}{\operatorname{det}(1-\hat{H}^{-})}\right) \\
    &-\frac{1}{2}\int_{-\infty}^{+\infty}dv G(-v-\ri\xi)\ln(1+e^{\ri Z(v+\ri\xi)})  +\frac{1}{2}\ln(1+e^{\ri Z(0)})  .  
\end{split}
\end{align}

\section{Bulk and boundary energy}
\label{app:bulk}
In this appendix, we present the derivation for our claim \eqref{eq:boundaryEe} in the main text. Strictly speaking, this is not really necessary for the derivation of the $g$-function, but we present it here as a consistency check of the lattice approach.
\subsection{Bulk Energy}

The bulk energy of the ground state comes from the eigenvalue of the operator $\widehat{\tau}(\Theta;\{\pm\Theta\})\tau(\Theta;\{\pm\Theta\}) $ corresponding to the AFV state, given by
\begin{align*}
   \tau_R(-2\Theta-\ri\gamma)= \left[\sinh(2\Theta+\mathrm{i}\gamma)\sinh(\mathrm{i}\gamma)\right]^{2N}
    \prod_{k=1}^N \left(\frac{\sinh(u_k-\Theta+\tfrac{\mathrm{i}\gamma}{2})}{\sinh(u_k-\Theta-\tfrac{\mathrm{i}\gamma}{2})}\frac{\sinh(u_k+\Theta-\tfrac{\mathrm{i}\gamma}{2})}{\sinh(u_k+\Theta+\tfrac{\mathrm{i}\gamma}{2})} \right) ,  
\end{align*}
Following the same steps as in Sec.~\ref{app:deriveG}, we have  
\begin{align}
\label{app:bulkE1}
    &\ln\prod_{k=1}^{N}\frac{\sinh(u_k-\Theta+\tfrac{\ri\gamma}{2})}{\sinh(u_k-\Theta-\tfrac{\ri\gamma}{2})}  \notag\\
    =&\ri\frac{N\pi}{\gamma}\int_{-\infty}^{+\infty}\frac{dv}{2\pi}\phi_{1/2}(\Theta-v)\left(\operatorname{sech}\left(\tfrac{\pi}{\gamma}(v+\Theta)\right)+\operatorname{sech}\left(\tfrac{\pi}{\gamma}(v-\Theta)\right)\right)   \notag\\
    &-\int_{-\infty}^{+\infty}\frac{dx\, dv}{2\pi}\phi_{1/2}'(\Theta-v-\ri\xi)(\delta-G)(v-x)\ln(1+e^{\ri Z_N(x+\ri\xi)})  \notag\\
    &+\int_{-\infty}^{+\infty}\frac{dx\, dv}{2\pi}\phi_{1/2}'(\Theta-v+\ri\xi)(\delta-G)(v-x)\ln(1+e^{-\ri Z_N(x-\ri\xi)})\,.
\end{align}
and
\begin{align}
\label{app:bulkE2}
    &\ln\prod_{k=1}^{N}\frac{\sinh(u_k+\Theta-\tfrac{\ri\gamma}{2})}{\sinh(u_k+\Theta+\tfrac{\ri\gamma}{2})}  \notag\\
    =&\ri\frac{N\pi}{\gamma}\int_{-\infty}^{+\infty}\frac{dv}{2\pi}\phi_{1/2}(\Theta+v)\left(\operatorname{sech}\left(\tfrac{\pi}{\gamma}(v+\Theta)\right)+\operatorname{sech}\left(\tfrac{\pi}{\gamma}(v-\Theta)\right)\right)   \notag\\
    &+\int_{-\infty}^{+\infty}\frac{dx\, dv}{2\pi}\phi_{1/2}'(\Theta+v+\ri\xi)(\delta-G)(v-x)\ln(1+e^{\ri Z_N(x+\ri\xi)})  \notag\\
    &-\int_{-\infty}^{+\infty}\frac{dx\, dv}{2\pi}\phi_{1/2}'(\Theta+v-\ri\xi)(\delta-G)(v-x)\ln(1+e^{-\ri Z_N(x-\ri\xi)}) .  
\end{align}
The energy of the AFV state is given by
\begin{align}
    E_0 \sim -\frac{\ri}{\Delta} \ln \prod_{k=1}^N \left(\frac{\sinh(u_k-\Theta+\tfrac{\mathrm{i}\gamma}{2})}{\sinh(u_k-\Theta-\tfrac{\mathrm{i}\gamma}{2})}\frac{\sinh(u_k+\Theta-\tfrac{\mathrm{i}\gamma}{2})}{\sinh(u_k+\Theta+\tfrac{\mathrm{i}\gamma}{2})} \right) .  
\end{align}
Since in \eqref{app:bulkE1} and \eqref{app:bulkE2}, only the first term on the right hand side can contribute to the order $\mathcal{O}(m R)$ in the field theory limit, we will focus on these terms. Taking their sum, we obtain 
\begin{align}
    &-\frac{N^2 \pi}{\gamma  R}\int_{-\infty}^{+\infty}\frac{dv}{2\pi}\left( \phi_{1/2}(\Theta+v) + \phi_{1/2}(\Theta-v)\right)\left(\operatorname{sech}\left(\tfrac{\pi}{\gamma}(v+\Theta)\right)+\operatorname{sech}\left(\tfrac{\pi}{\gamma}(v-\Theta)\right)\right) \notag\\
    =&\frac{2\ri N^2}{R} \int_{-\infty}^{\infty} dk \frac{e^{\ri k\Theta}\sinh\left(\tfrac{k}{2}(\pi-\gamma)\right)\cos(k\Theta)}{k\sinh\left(\tfrac{\pi k}{2}\right)\cosh\left(\tfrac{\gamma k}{2}\right)} 
\end{align}
The integral as it stands is divergent. This is understandable physically because it contains UV divergences. To extract meaningful results in the field theory limit, one should extract the order $\mathcal{O}\left(\exp\left(-\tfrac{2\pi\Theta}{\gamma}\right)\right)$  contribution of the integral, which comes from the contribution of the pole at $k=\tfrac{\ri\pi}{\gamma}$. In the field theory limit, the contribution of this pole gives the bulk energy 
\begin{align}
   E_0 \equiv R\mathcal{E}= \frac{m^2 R}{4} \cot\left(\frac{\pi^2}{2\gamma}\right) .  
\end{align}
$\mathcal{E}$ is precisely the energy density for the vacuum state of sine-Gordon theory known in the literature \cite{Destri:1994bv}.

\subsection{Boundary Energy}
\label{app:BoundaryEnergy2}
The ground state boundary energy $\varepsilon_a$ is encoded in the overlap as
\begin{align}
\label{app:bdyenergy}
   g_a g_b^* e^{-R(\varepsilon_{a}+\varepsilon_{b})} = \langle B_a | 0 \rangle \langle 0 | B_b \rangle .  
\end{align}
For simplicity, we will consider the case $\varepsilon_a=\varepsilon_b$. From the discussions in appendix \ref{app:gfunc}, the boundary energy is encoded in the first term on the right hand side of \eqref{app:prefact0}, whose form we quote here
\begin{align}
\label{eq:divterm}
    \frac{N}{2\gamma} \int_{-\infty}^{+\infty} dx \ln f(-x+\ri\xi)  \left(\operatorname{sech}\left(\tfrac{\pi}{\gamma}(x+\Theta-\ri\xi)\right)+\operatorname{sech}\left(\tfrac{\pi}{\gamma}(x-\Theta-\ri\xi)\right)\right) \,,  
\end{align}
 which is divergent in the continuum limit. To extract meaningful results, we focus on the contributions of order $\mathcal{O}\left(\exp(-\frac{\pi\Theta}{\gamma})\right)$ in the integral such that in the continuum limit after combining with the $N$ in front of the integral, it gives finite result by the following relationship 
\begin{align}
    \frac{m R}{4}=N\exp(-\frac{\pi\Theta}{\gamma}) 
\end{align}
All terms of orders higher than $\mathcal{O}\left(\exp(-\frac{\pi\Theta}{\gamma})\right)$ should corresponds to UV divergences in the continuum limit. 

To select the terms of correct order, one can write the integral in the Fourier space. Using contour integral, for $0<\xi<\tfrac{\gamma}{2}$, we have
\begin{align} 
    I^{\pm} \equiv \mathcal{F}\left\{ \text{sech} \left( \tfrac{\pi}{\gamma}(x\pm\Theta-\ri\xi) \right) \right\}(k) = \gamma \exp\left( -\ri k(\ri\xi\mp\Theta) \right) \text{sech} \left( \tfrac{\gamma k}{2} \right). 
\end{align} 
Thus 
\begin{align}
    I^{+}+I^{-} = 2\gamma e^{k\xi} \text{sech} \left( \tfrac{\gamma k}{2} \right) \cos\left(k\Theta\right).   
\end{align}
To deal with the logarithm in \eqref{eq:divterm}, we can take the derivative with respect to $u$, 
\begin{align}
\label{eq:fourterm}
    \frac{d}{du}\ln f(u+\ri\xi)=2 \left(\coth(u+\ri a+\ri\xi)+\tanh(u+b+\ri\xi)-\coth(2u+\ri\gamma+2\ri\xi)-\coth(2u+2\ri\xi)\right),   
\end{align}
where $a$ and $b$ are real and we consider the case $0<a<\pi$ for convenience. 

There are four terms on the right hand side of \eqref{eq:fourterm}. Their integration can be computed similarly. We discuss the first term in detail. For $0<a+\xi<\pi$, we have
\begin{align}
    \int_{-\infty}^{+\infty} e^{-\ri ku} \text{coth}(u+\ri a+\ri\xi) du = -\ri\pi \frac{e^{k\left(\tfrac{\pi}{2}-a-\xi\right)}}{\text{sinh}\left(\tfrac{\pi k}{2}\right)}.  
\end{align}
Thus 
\begin{align}
\label{eq:interTerm1}
    &\int_{-\infty}^{+\infty} \text{coth}(u-x+\ri a+\ri\xi) \left(\operatorname{sech}\left(\tfrac{\pi}{\gamma}(x+\Theta-\ri\xi)\right)+\operatorname{sech}\left(\tfrac{\pi}{\gamma}(x-\Theta-\ri\xi)\right)\right) dx \notag\\
    =&\frac{1}{2\pi} \int_{-\infty}^{+\infty} e^{\ri ku} (-2i\pi\gamma) \frac{e^{k(\tfrac{\pi}{2}-a)}\cos(k\Theta)}{\sinh(\tfrac{\pi k}{2})\cosh(\tfrac{\gamma k}{2})} dk .  
\end{align}
After integrating the rhs of \eqref{eq:interTerm1} over $u$, we have
\begin{align}
\label{eq:tempInt1}
    \frac{-\ri}{2\pi} \int_{-\infty}^{+\infty} e^{\ri ku} (-2\ri\pi\gamma) \frac{e^{k(\tfrac{\pi}{2}-a)}\cos(k\Theta)}{k\sinh(\tfrac{\pi k}{2})\cosh(\tfrac{\gamma k}{2})}dk.  
\end{align}
The $\mathcal{O}\left(\exp(-\frac{\pi\Theta}{\gamma})\right)$ contribution is given by the residues at the poles $\tfrac{\ri\pi}{\gamma}$ and $-\tfrac{\ri\pi}{\gamma}$. The contribution of the pole $\tfrac{\ri\pi}{\gamma}$ is (taking $u=0$)
\begin{align}
    4\gamma e^{\tfrac{\ri\pi}{2\gamma}(\pi-2a)} \text{csc}\left(\frac{\pi^2}{2\gamma}\right) \cosh\left(\frac{\pi\Theta}{\gamma}\right)
\end{align}
and the contribution of the pole $-\tfrac{\ri\pi}{\gamma}$ is (taking $u=0$)
\begin{align}
    4\gamma e^{-\tfrac{\ri\pi}{2\gamma}(\pi-2a)} \text{csc}\left(\frac{\pi^2}{2\gamma}\right) \cosh\left(\frac{\pi\Theta}{\gamma}\right). 
\end{align}
The summation gives
\begin{align}
    8\gamma \cos\left(\tfrac{\ri\pi}{2\gamma}(\pi-2a)\right) \text{csc}\left(\frac{\pi^2}{2\gamma}\right) \cosh\left(\frac{\pi\Theta}{\gamma}\right). 
\end{align}
Multiplying by $\tfrac{N}{2\gamma}\cdot2$ and taking the continuum limit, we have
\begin{align}
\label{eq:DTerm1}
    mR\frac{\cos\left(\tfrac{\nu}{2}(\pi-2a)\right)}{\sin\left(\tfrac{\pi\nu}{2}\right)},   
\end{align}
where we used $\nu=\frac{\pi}{\gamma}$.   

For $0<\xi<\frac{\pi}{2}$, the order $\mathcal{O}\left(\exp(-\frac{\pi\Theta}{\gamma})\right)$ contribution of the second term 
\begin{align}
    \frac{N}{2\gamma}\int_{-\infty}^{+\infty} 2\ln\text{cosh}(-x+b+\ri\xi) \left(\operatorname{sech}\left(\tfrac{\pi}{\gamma}(x+\Theta-\ri\xi)\right)+\operatorname{sech}\left(\tfrac{\pi}{\gamma}(x-\Theta-\ri\xi)\right)\right) dx    
\end{align}
is given by 
\begin{align}
\label{eq:DTerm2}
    m R \frac{\cosh(\nu b)}{\sin\left(\tfrac{\pi\nu}{2}\right)}.  
\end{align}

For $0<\tfrac{\gamma}{2}+\xi<\tfrac{\pi}{2}$, the order $\mathcal{O}\left(\exp(-\frac{\pi\Theta}{\gamma})\right)$ contribution of the third term
\begin{align}
    -\frac{N}{2 \gamma}\int_{-\infty}^{+\infty} \ln\text{sinh}(-2x+\ri\gamma+2\ri\xi) \left(\operatorname{sech}\left(\tfrac{\pi}{\gamma}(x+\Theta-\ri\xi)\right)+\operatorname{sech}\left(\tfrac{\pi}{\gamma}(x-\Theta-\ri\xi)\right)\right) dx 
\end{align}
is given by 
\begin{align}
\label{eq:DTerm3}
    -\frac{m R}{2}.  
\end{align}

For $0<\xi<\tfrac{\pi}{2}$, the order $\mathcal{O}\left(\exp(-\frac{\pi\Theta}{\gamma})\right)$ contribution of the fourth term  
\begin{align}
    -\frac{N}{2 \gamma}\int_{-\infty}^{+\infty} \ln\text{sinh}(-2x+2\ri\xi) \left(\operatorname{sech}\left(\tfrac{\pi}{\gamma}(x+\Theta-\ri\xi)\right)+\operatorname{sech}\left(\tfrac{\pi}{\gamma}(x-\Theta-\ri\xi)\right)\right) dx  
\end{align}
is given by  
\begin{align}
\label{eq:DTerm4}
    -\tfrac{mR}{2}\text{cot}  \left(\tfrac{\pi\nu}{4}\right) .  
\end{align}

Summing the four terms \eqref{eq:DTerm1}, \eqref{eq:DTerm2}, \eqref{eq:DTerm3} \eqref{eq:DTerm4}, we find that in the field theory limit, after subtracting UV divergence, the order $\mathcal{O}(mR)$ contribution from $\ln\prod_{j=1}^{N}f(u_j)$ is given by
\begin{align*}
    \tfrac{mR}{2} \left(-\text{cot}  \left(\tfrac{\pi\nu}{4}\right)-1+2  \tfrac{\cos\left(\tfrac{\nu}{2}(\pi-2a)\right)}{\sin\left(\tfrac{\pi\nu}{2}\right)} +2  \tfrac{\cosh(\nu b)}{\sin\left(\tfrac{\pi\nu}{2}\right)} \right) . 
\end{align*}
Comparing with \eqref{app:bdyenergy}, we find the boundary energy 
\begin{align}
\label{eq:boundaryEfinal}
    2 \varepsilon_{a} = - \tfrac{m}{2} \left(-\text{cot}  \left(\tfrac{\pi\nu}{4}\right)-1+2  \tfrac{\cos\left(\tfrac{\nu}{2}(\pi-2a)\right)}{\sin\left(\tfrac{\pi\nu}{2}\right)} +2  \tfrac{\cosh(\nu b)}{\sin\left(\tfrac{\pi\nu}{2}\right)} \right) .  
\end{align}
Let us comment on a few subtle points on the computation of boundary energy. First, the result matches that given in the literature \cite{Ahn:2003ns} for $a>\tfrac{\gamma}{2}$. Second, in computing \eqref{eq:tempInt1} and other similar integrals, we pick up two poles at $\pm\tfrac{\ri \pi}{\gamma}$ since they contribute to finite results in the continuum limit. However, how to deform the contour from the real axis to pick up precisely these two poles is not completely clear. If we pick only one of the poles, the final result for $\varepsilon_a+\varepsilon_b$ would be slightly different from \eqref{eq:boundaryEfinal}. It would be nice to have a better understanding on these points. Nevertheless, they do not affect our result for the $g$-function.

\section{Free Fermion Point}
\label{app:FreeF}
The sine-Gordon theory at bulk parameter $\gamma=\tfrac{\pi}{2}$ is equivalent to the free massive fermion by bosonization \cite{Ameduri:1995vh}. The free fermion point is particularly simple and can serve as a consistency check. In this appendix, we compute the $g$-function at the free fermion point with the fixed boundary condition by TBA and show that the result is equivalent to the lattice regularization approach.  

\subsection{Thermodynamic Bethe Ansatz Approach to $g$-function}

At the free fermion point, we have solitons and anti-solitons which scatter freely in the bulk. The boundary scattering process of solitons and anti-solitons can be described by the following Faddeev-Zamolodchikov (FZ) algebra \cite{Ameduri:1995vh}
\begin{align}
    & A_{+}^{\dagger}(\theta) B=P^{+}(\theta) A_{+}^{\dagger}(-\theta) B+Q^{+}(\theta) A_{-}^{\dagger}(-\theta) B\,, \\
    & A_{-}^{\dagger}(\theta) B=Q^{-}(\theta) A_{+}^{\dagger}(-\theta) B+P^{-}(\theta) A_{-}^{\dagger}(-\theta) B\,,  
\end{align}
where $A^{\dagger}_+$, $A^{\dagger}_-$ and $B$ are FZ operators for soliton, anti-soliton and the boundary respectively. The free fermion theory with fixed boundary condition corresponds to the case $Q^{+}(\theta)=Q^{-}(\theta)=0$ and
\begin{align}
    P^{\pm}(\theta)=-\left.\cosh \left(\frac{\theta \pm \ri \sqrt{4\pi}\varphi_0 - \ri \pi / 2}{2}\right) \right/ \cosh \left(\frac{\theta \mp \ri \sqrt{4\pi}\varphi_0+\ri \pi / 2}{2}\right).  
\end{align}
Note that we take boundary parameters of the two boundaries to be the same for convenience. The boundary parameter here is related to lattice parameter by $\sqrt{4\pi}\varphi_0+2\pi n=-2a$ when $\lambda=1$ \cite{Ameduri:1995vh,Ahn:2003ns}.  Using the relation between $\varphi_0$ and $a$, we have
\begin{align}
    P^{\pm}(\theta)=-\left.\cosh\left(\frac{\theta-\ri\pi/2}{2}\mp  \ri a\right)\right/ \cosh\left(\frac{\theta+\ri\pi/2}{2}\pm  ia\right) . 
\end{align}

Assuming that there are $N^{+}$ fermions and $N^{-}$ anti-fermions, the Bethe-Yang equations are
\begin{align}
    &e^{\ri2LM\sinh\theta_j}(P^{+}(\theta_j))^{2}=1, \qquad j=1,\dots,N^{+} \\
    &e^{\ri2LM\sinh\theta_j}(P^{-}(\theta_j))^{2}=1, \qquad j=1,\dots,N^{-}. 
\end{align}
In the thermodynamic limit, they become
\begin{align}
    &\rho_{\text{tot}}^{+}=\frac{m}{\pi}\cosh\theta+\frac{1}{2\pi iL}\partial_{\theta}\ln (P^{+}(\theta))^{2}-\frac{\delta(\theta)}{2L}, \\
    &\rho_{\text{tot}}^{-}=\frac{m}{\pi}\cosh\theta+\frac{1}{2\pi iL}\partial_{\theta}\ln (P^{-}(\theta))^{2}-\frac{\delta(\theta)}{2L}.  
\end{align}
Defining $Y$-functions as 
\begin{align}
    Y_{\pm}=\frac{\rho^{\pm}}{\rho_{\text{tot}}^{\pm}-\rho^{\pm}}\,,
\end{align}
the TBA equations read
\begin{align}
    \ln Y_{\pm}=-Rm\cosh(\theta). 
\end{align}
The $g$-function is given by the $Y$-functions as
\begin{align}
    \ln |g|^2_{\text{(TBA)}}=2\sum_{A=\pm}\int_{0}^{\infty}\frac{d\theta}{2\pi}\Theta_A(\theta)\ln(1+Y_A(\theta))
\end{align}
where 
\begin{align}
    \label{chap3.34}
    \Theta_A(\theta)\equiv \frac{1}{\ri}\partial_{\theta}\ln P^{A}(\theta)- \pi \delta(\theta) . 
\end{align}
There is no Fredholm determinant factor in $g$-function because the S-matrix of free fermion is trivial. Using the explicit expression of the reflection factor, we have
\begin{align}
    \frac{1}{\ri}\partial_{\theta}\ln P^{\pm}(\theta)=\frac{\ri}{2}\left(\tanh(\pm  \ri a\mp\tfrac{\theta}{2}+\tfrac{\ri\pi}{4})+\tanh(\pm  \ri a\pm\tfrac{\theta}{2}+\tfrac{\ri\pi}{4})\right). 
\end{align}

\subsection{NLIE Approach to $g$-function}
Now we compute the $g$-function using the lattice regularization approach. We take $\gamma=\tfrac{\pi}{2}$ in the NLIE and the expressions \eqref{eq:appC2gfunc} for the $g$-function . In this case, $G(u)=0$ and the NLIE simplifies to 
\begin{align}
    Z(u)=mR\sinh\left(\frac{\pi u}{\gamma}\right)=mR\sinh(2 u).  
\end{align}
As a result,
\begin{align}
    \ri Z(u+\tfrac{\ri\gamma}{2})=-mR\cosh(2u), \\
    -\ri Z(u-\tfrac{\ri\gamma}{2})=-mR\cosh(2u). 
\end{align} 
Taking the limit $\xi\to\tfrac{\pi}{4}$ and assuming that $\tfrac{\pi}{4}<a<\tfrac{3\pi}{4}$, the $g$-function becomes 
\begin{align}
\label{eq:ddvGappendix}
    \ln |g|^2_{\text{(NLIE)}}=-\frac{1}{\pi} \operatorname{Im}\left\{ \int_{-\infty}^{+\infty} dv  (\ln f_{D})'(-v-\ri\frac{\pi}{4})   \ln(1+e^{-mR\cosh(2 v )}) \right\}  -  \ln 2 
\end{align}
Note that since we are considering the diagonal boundary scattering, we should take the following diagonal limit for the function $f(u)$ 
\begin{align}
    f_{D}(u)\equiv \lim_{b\xrightarrow{}\infty}e^{-2b}f(u)=\frac{\sinh^2(u+\ri a) e^{2u}}{\sinh(2u+\ri\gamma)\sinh(2u)}.  
\end{align}
It follows that 
\begin{align}
\label{eq:logfD}
    (\ln f_D)'(-v-\ri\tfrac{\pi}{4})=2\tanh(\ri a-v+\tfrac{\ri\pi}{4}) + 2\tanh(2v) + 2\coth(2v-\ri\varepsilon) + 2 ,   
\end{align}
where $0<\varepsilon\ll 1$ is introduced to avoid the singularity on the integration contour. Inside the integral \eqref{eq:ddvGappendix}, equation \eqref{eq:logfD} can be rewritten as 
\begin{align}
    (\ln f_D)'(-v-\ri\tfrac{\pi}{4})=2\tanh(\ri a-v+\tfrac{\ri\pi}{4}) + 2\tanh(2v) + 2 \mathcal{P}\coth(2v) + 2 \pi \ri \delta(2 v) + 2 ,     
\end{align}
where $\mathcal{P}$ means the principal value. It's easy to check that
\begin{align}
    &-\frac{1}{\pi} \operatorname{Im}\left\{ \int_{-\infty}^{+\infty} dv  (\ln f_{D})'(-v-\ri\frac{\pi}{4})   \ln(1+e^{-mR\cosh(2 v )}) \right\} \notag\\
    =& 2\sum_{A=\pm}\int_{0}^{\infty}\frac{d\theta}{2\pi}\Theta_A(\theta) \ln(1+Y_A(\theta)) .  
\end{align}
Then we have 
\begin{align}
    \ln |g|^2_{\text{(NLIE)}} = \ln |g|^2_{\text{(TBA)}} - \ln 2 . 
\end{align}
Note that when the boundary parameter $a$ goes beyond the regime $\tfrac{\pi}{4}<a<\tfrac{3\pi}{4}$, there will be extra terms in the expression \eqref{eq:ddvGappendix}. This may corresponds to the fact that if we go to the regime $\zeta>\pi/2$ or $\zeta<-\pi/2$ in the Dirichlet limit, certain boundary bound states will show up in the physical strip \cite{Ghoshal:1993tm,Bajnok:2001ug}. This deserves a further more detailed analysis, which we leave for future work.

\section{IR and UV limits of the $g$-function}
\label{app:IRUV}
In this appendix we compute the IR and UV limits of the $g$-function for the sine-Gordon theory.

\subsection{IR Limit}
We first consider the IR limit $mR \gg 1$. Following the procedure in \cite{Destri:1994bv}, we define the following quantities, 
\begin{align}
    &\epsilon(\theta)\equiv -\ri Z(\tfrac{\gamma}{\pi}\theta+\ri\tfrac{\gamma}{2}) , \\
    &\bar{\epsilon}(\theta)\equiv \ri Z(\tfrac{\gamma}{\pi}\theta-\ri\tfrac{\gamma}{2}) , \\
    &G_{0}(\theta)\equiv \tfrac{\gamma}{\pi}G(\tfrac{\gamma}{\pi}\theta)=\int_{-\infty}^{+\infty}\frac{dk}{4\pi}e^{\ri k\theta}\frac{\sinh\left((\tfrac{\pi^2}{2\gamma}-\pi)k\right)}{\sinh\left((\tfrac{\pi^2}{2\gamma}-\tfrac{\pi}{2})k\right)\cosh\left(\tfrac{\pi}{2}k\right)}
\end{align}
In terms of these quantities, NLIE becomes 
\begin{align}
\label{eq:NLIEtba}
    \epsilon(\theta)=mR\cosh(\theta)&-\int_{-\infty}^{+\infty}dv\, G_{0}(\theta-v)\ln(1+e^{-\epsilon(v)})  \notag\\ 
    &+\int_{-\infty}^{+\infty}dv\, G_{0}(\theta-v+i\pi-i0^{+})\ln(1+e^{-\bar{\epsilon}(v)})  .
\end{align}
In the IR limit, the function $\epsilon(\theta)$ can be expanded in the following way 
\begin{align}
    \epsilon(\theta)=f_{0}(\theta)+f_{1}(\theta)+f_{2}(\theta)+\dots
\end{align}
where each term has one more $e^{-mR}$ factor comparing with its last one. The first term in this expansion can be obtained by ignoring the integral terms in the functional equation \eqref{eq:NLIEtba} and the expression for other terms can be obtained by putting the expansion into the functional equation \eqref{eq:NLIEtba}, \emph{i.e.}, $\epsilon(\theta)\sim mR\cosh(\theta)$ in the IR limit. 

To treat the determinant part in the IR limit, note that one may choose the contour $\mathcal{C}$ in \eqref{eq:DetExpan} and \eqref{eq:Hoperat} to be the union of $\mathbb{R}+\tfrac{\ri\gamma}{2}-\ri\varepsilon$ and $\mathbb{R}-\tfrac{\ri\gamma}{2}+\ri\varepsilon$, where the small $\varepsilon$ is introduced to avoid the possible singularities in the integrand. It is useful to expand the following factors in the IR limit, 
\begin{align}
\label{eq:detafactorp}
    &\frac{\mathfrak{a}(u+\frac{\ri\gamma}{2})}{1+\mathfrak{a}(u+\tfrac{\ri\gamma}{2})}=e^{\ri Z((u+\frac{\ri\gamma}{2}))}-e^{2\ri Z((u+\frac{\ri\gamma}{2}))}+\mathcal{O}(e^{-3mR}),  \\
\label{eq:detafactorm}
    &\frac{\mathfrak{a}(u-\frac{\ri\gamma}{2})}{1+\mathfrak{a}(u-\frac{\ri\gamma}{2})}=1-e^{-\ri Z((u-\frac{\ri\gamma}{2}))}+\mathcal{O}(e^{-2mR}), \\  
\label{eq:limitlog1pa}
    &\ln(1+e^{\ri Z(u+\frac{\ri\gamma}{2})})\sim e^{\ri Z(u+\frac{\ri\gamma}{2})} + \mathcal{O}(e^{-2mR}).  
\end{align}

From the equation \eqref{eq:DetExpan} we know that the leading contribution of the logarithm of the $g$-function should be of order $\mathcal{O}(1)$. Only the integral along $\mathbb{R}-\tfrac{\ri\gamma}{2}$ will contribute at this order. A general term that contributes in the expansion includes the expression of the form 
\begin{align}
\begin{split}
    &\int_{-\infty}^{\infty}\frac{dz_1\dots dz_n}{(2\pi \ri)^n}\varphi(z_1+z_2-\ri\gamma) \varphi(z_2-z_3) \dots \varphi(z_{n-1}-z_{n}) \varphi(z_n-z_1) \\
    =&\int_{-\infty}^{\infty}\frac{dz_1\dots dz_n}{(2\pi \ri)^n} \varphi(2z_1-\ri\gamma) \varphi(z_2) \varphi(z_3) \dots \varphi(z_n) ,
\end{split}
\end{align}
where $\varphi(u)$ is given in \eqref{eq:varphi}. Thus to evaluate the quantity \eqref{eq:DetExpan} at the leading order in the IR limit, one should evaluate the following integrals: 
\begin{align}
\label{eq:IRmidint1}
    &\textbf{I}\equiv   \int_{-\infty+\ri\varepsilon}^{\infty+\ri\varepsilon}\frac{dz}{2\pi \ri}\left( \frac{-\sinh(2\ri\gamma)}{\sinh(2z)\sinh(2z-2i\gamma)} \right)=\frac{1}{2\pi \ri}\left(\frac{\ri\pi}{2}+2\ri\gamma\right)=\frac{\gamma}{\pi} - \frac{1}{2}  \\
\label{eq:IRmidint2}
    &\textbf{II}\equiv   \int_{-\infty}^{\infty}\frac{dz}{2\pi \ri}\left( \frac{-\sinh(2\ri\gamma)}{\sinh(z+\ri\gamma)\sinh(z-\ri\gamma)} \right)=\frac{1}{2\pi \ri}(-2\ri\pi+4\ri\gamma)= \frac{2\gamma}{\pi}-1 . 
\end{align}
As we mentioned before, the small $\ri\varepsilon$ in the \eqref{eq:IRmidint1} means that the integral should be done slightly above the real axis to avoid possible singularity.  

Putting the results into the equation \eqref{eq:DetExpan}, we find that 
\begin{align}
    \ln\frac{\operatorname{det}(1-\hat{H}^{+})}{\operatorname{det}(1-\hat{H}^{-})} = -\sum_{n=1}^{\infty}\frac{1}{n}\textbf{I} \cdot\textbf{II}^{n-1} =\frac{\textbf{I}}{\textbf{II}}\ln(1-\textbf{II})=\frac{1}{2}\ln\left(2-\frac{2\gamma}{\pi}\right)  .  
\end{align}
Assuming that $0<\gamma<\pi/2$, $a$ and $b$ are real with $0<a<\pi-\tfrac{\gamma}{2}$, the leading order contribution should be
\begin{align}
    \ln|g|^2 \sim \frac{\textbf{I}}{\textbf{II}}\ln(1-\textbf{II})=\frac{1}{2}\ln\left(2-\frac{2\gamma}{\pi}\right)-\ln(1+e^{iZ(0)}) = \frac{1}{2}\ln\left(\frac{1}{2}-\frac{\gamma}{2\pi}\right) ,   
\end{align}
where we used $Z(0)=0$. This comes from the definition of $Z$ function at the lattice level. For the free boundary condition, where $a=b=0$, the leading order contribution in the IR limit should be 
\begin{align}
    \ln|g|^2 \sim \frac{1}{2}\ln\left(2-\frac{2\gamma}{\pi}\right) + \ln(1+e^{iZ(0)}) = \frac{1}{2}\ln \left[ 8\left( 1-\frac{\gamma}{\pi}\right) \right] .  
\end{align}

\subsection{UV Limit}
Now we consider the UV limit $mR \rightarrow 0$. We will argue that the logarithm of $g$-function should be zero in this limit. First, the logarithm of the Fredholm determinant part should be zero in the UV limit. Take the integral contour in \eqref{eq:DetExpan} to be $\{\mathbb{R}+\ri\xi\}\cup\{\mathbb{R}-\ri\xi\}$, with $0<\xi<\tfrac{\gamma}{2}$, and take $\xi$ to be sufficiently small such that no singularity coming from the function $\varphi(z)$ is included. Numerical evidence shows that in the UV limit, the solution of the non-linear integral equation is $\mathfrak{a}(z\pm \ri\xi)=1$, where $z$ is real. For convenience, introduce the following notations 
\begin{align}
    \textbf{A} \equiv \frac{\mathfrak{a}\left(z-\ri\xi\right)}{1+\mathfrak{a}\left(z-\ri\xi\right)} ,\qquad \textbf{B} \equiv \frac{\mathfrak{a}\left(z+\ri\xi\right)}{1+\mathfrak{a}\left(z+\ri\xi\right)} .
\end{align}
Then the $n$-th term in the expansion \eqref{eq:DetExpan} becomes 
\begin{align}
    -\frac{\mathbf{I}\cdot\mathbf{II}^{n-1}}{n} \left( \textbf{A}^n - C_{n}^{1} \textbf{A}^{n-1}\textbf{B} + C_{n}^{2} \textbf{A}^{n-2}\textbf{B}^2 -\dots + (-1)^n \textbf{B}^n \right) = -\frac{\mathbf{I}\cdot\mathbf{II}^{n-1}}{n} (\textbf{A}-\textbf{B})^n = 0 ,  
\end{align}
where we have used the fact that $\textbf{A}=\textbf{B}=\tfrac{1}{2}$ for $z\in\mathbb{R}$.  

Second, the following logarithm of prefactor part 
\begin{align}
    &-\frac{1}{2\pi} \operatorname{Im}\left\{ \int_{-\infty}^{+\infty} dv \, dx (\ln f)'(-x-i\xi)  \,  (\delta-G)(x-v) \ln(1+e^{iZ(v+i\xi)}) \right\} \notag\\
    &+\frac{1}{2}\int_{-\infty}^{+\infty}dv G(-v-i\xi)\ln(1+e^{iZ(v+i\xi)})  -\frac{1}{2}\ln(1+e^{iZ(0)}) 
\end{align}
becomes
\begin{align}
    \frac{1}{2} \frac{1}{2\pi i} \ln 2 \left[ \int_{-\infty}^{\infty}  (\delta-G)(-v) dv \int_{\mathcal{C}} (\ln f)'(-z) dz \right] + \frac{1}{2} \ln 2 \int_{-\infty}^{\infty} G(-v) dv -\frac{1}{2} \ln 2    
\end{align}
in the UV limit. Since
\begin{align}
    \int_{\mathcal{C}} (\ln f)'(-z) dz = 2\pi \ri ,  
\end{align}
the logarithm of the prefactor part is also vanishing in the UV limit.

\section{Numerical calculations}
\label{app:num}
In this appendix, we give details for the numerical method for solving NLIE and computing the Fredholm determinants.

\subsection{Solving the Non-linear Integral Equation}
The starting point is the following non-linear integral equation for vacuum: 
\begin{align}
\label{app:nliesol}
    Z(u)=mR\sinh\left(\frac{\pi u}{\gamma}\right)+2\operatorname{Im}\int_{-\infty}^{+\infty}\mathrm{d}v  G(u-v-\ri\xi)\ln(1+e^{\ri Z(v+\ri\xi)})\,.
\end{align}
This equation can be analytically continued and rewritten as 
\begin{align}
\label{eq:NLIE1}
\begin{split}
    \epsilon(\theta,\xi)=-\ri mR\sinh(\theta+\ri\xi)&-\int_{-\infty}^{+\infty}\mathrm{d}v G_{0}(\theta-v)\ln(1+e^{-\epsilon(\theta,\xi)})+\int_{-\infty}^{+\infty}\mathrm{d}v G_{1}(\theta-v)\ln(1+e^{-\bar{\epsilon}(\theta,\xi)})
\end{split}
\end{align}
where
\begin{align}
    \epsilon(\theta,\xi)=-\ri Z(\tfrac{\gamma}{\pi}\theta+\ri\tfrac{\gamma}{\pi}\xi)
\end{align}
and
\begin{align}
    G_{0}(\theta)&=\int_{-\infty}^{+\infty}\frac{\mathrm{d}k}{2\pi}e^{-\ri k\theta}\frac{\sinh((\tfrac{\pi^2}{2\gamma}-\pi)k)}{2\sinh((\frac{\pi^2}{2\gamma}-\frac{\pi}{2})k)\cosh(\tfrac{\pi k}{2})}\equiv \int_{-\infty}^{+\infty}e^{-\ri k\theta}Y(k)\frac{\mathrm{d}k}{2\pi}, \\
    \label{eq:G1}
    G_{1}(\theta,\xi)&=\int_{-\infty}^{+\infty}\frac{\mathrm{d}k}{2\pi}e^{-\ri k\theta}\frac{e^{2k\xi}\sinh((\tfrac{\pi^2}{2\gamma}-\pi)k)}{2\sinh((\frac{\pi^2}{2\gamma}-\frac{\pi}{2})k)\cosh(\tfrac{\pi k}{2})}\equiv \int_{-\infty}^{+\infty}e^{-\ri k\theta}e^{2k\xi}Y(k)\frac{\mathrm{d}k}{2\pi}.  
\end{align}
To solve this equation numerically, we take the following steps. 
\begin{enumerate}
    \item Make the grid. Give a cutoff $\texttt{L}$ to functions in the equation and divide the interval $[-\texttt{L},\texttt{L}]$ into $\texttt{N}$ part with length $\tfrac{2\texttt{L}}{\texttt{N}}$. 
    \item Sample the points in the grid on the variable $\theta$ of source function $-\ri mR\sinh(\theta+\ri\xi)$. 
    \item Calculate the convolution term using the convolution theorem. This can be done in the following way. Define the function $L(\theta,\xi)\equiv\ln(1+e^{-\epsilon(\theta,\xi)})$, then upon discretization, the convolution term can be written as
\begin{align}
\label{eq:conv0}
\begin{split}
    \int_{-\infty}^{+\infty}G_{0}(u-v)L(v)\mathrm{d}v
    \sim&\delta\sum_{m=-\mathtt{N}/2}^{\mathtt{N}/2-1}G_{0}(n\delta-m\delta)L(m\delta)  ,  
\end{split}
\end{align}
where $\delta=\tfrac{2\texttt{L}}{\texttt{N}}$. Note that we omit the $\xi$ dependence for convenience here. Define the series $\{G_n^{(0,1)}\equiv G_{0,1}(n\delta)\}$ and $\{L_{n}\equiv L(n\delta)\}$, where $n$ takes the integer from $-\mathtt{N}/2$ to $\mathtt{N}/2-1$. One can do periodic continuation for these two lists, \emph{i.e.}, set $G_{n+N}=G_{n}$ and $L_{n+N}=L_{n}$. Thus we can write the convolution term as
\begin{align}
    \int_{-\infty}^{+\infty}G_{0}(u-v)L(v)\mathrm{d}v \sim \delta\sum_{m=0}^{\texttt{N}-1}G_{0}(n\delta-m\delta)L(m\delta)   . 
\end{align}
To continue, one need the discretized $G_{0,1}$ function, which can be written as the discrete Fourier transformation of the discretized series of function $Y$. They are related in the following way 
\begin{align}
    G^{(0)}_{n}=&G_{0}(n\delta)=\int_{-\infty}^{+\infty}e^{-\ri kn\delta}Y(k)\frac{\mathrm{d}k}{2\pi}\sim\int_{-\pi/\delta}^{+\pi/\delta}e^{-\ri kn\delta}Y(k)\frac{\mathrm{d}k}{2\pi} = \frac{1}{\delta}\int_{-\pi}^{+\pi}e^{-\ri kn}Y(\frac{k}{\delta})\frac{\mathrm{d}k}{2\pi} \notag\\
    \sim & \frac{1}{2\texttt{L}}\sum_{m=-\texttt{N}/2}^{\texttt{N}/2-1}e^{-\ri2\pi mn/\texttt{N}}Y(\frac{\pi m}{\texttt{L}})\equiv\frac{1}{2\texttt{L}}\sum_{m=-\texttt{N}/2}^{\texttt{N}/2-1}e^{-\ri2\pi mn/\texttt{N}}Y_m =\frac{1}{2 \texttt{L}}\sum_{m=0}^{\texttt{N}-1} e^{-\ri2\pi mn/\texttt{N}} Y_m, 
\end{align}
where we also used the periodic continued $Y$ series, \emph{i.e.}, $Y_{m+N}=Y_{m}$. Similarly, 
\begin{align}
    G^{(1)}_{n}=&G_{0}(n\delta) \sim \frac{1}{2 \texttt{L}}\sum_{m=0}^{\texttt{N}-1} e^{-\ri2\pi mn/\texttt{N}} e^{2\xi\pi m/\texttt{L}}  Y_m .  
\end{align}
Then the convolution theorem gives that 
\begin{align}
    &\delta\sum_{n=0}^{\texttt{N}-1}\sum_{m=0}^{\texttt{N}-1}G_{0}(n\delta-m\delta)L(m\delta)e^{\ri2\pi pn/\texttt{N}} = Y_p\cdot\left(\sum_{m=0}^{\texttt{N}-1}L_m e^{\ri2\pi pm/\texttt{N}}\right) , \\
    &\delta\sum_{n=0}^{\texttt{N}-1}\sum_{m=0}^{\texttt{N}-1}G_{1}(n\delta-m\delta)L(m\delta)e^{\ri2\pi pn/\texttt{N}} = e^{2\xi\pi p/\texttt{L}} Y_p\cdot\left(\sum_{m=0}^{\texttt{N}-1}L_m e^{\ri2\pi pm/\texttt{N}}\right)   .  
\end{align}
The inverse Fourier transformation of the right hand side of these equations gives the convolution terms.  
    \item Sum all terms on the right hand side of the equation \eqref{eq:NLIE1} and set-up the iteration to solve the non-linear integral equation. Putting the source function $-\ri m\beta\sinh(\theta+\ri\xi)$ to be the initial function for iteration. After sufficient times of iteration, we can obtain the numerical solution of NLIE.          

\end{enumerate}

\subsection{Computation of Fredholm Determinant}
Recall the expression for the logarithm of $g$-function
\begin{align}
\label{app:gfunct}
    &- \operatorname{Im} \int_{-\infty}^{+\infty} \frac{dv \, dx}{2\pi} (\ln f)'(-x-\ri\xi)  \,  (\delta-G)(x-v) \ln(1+e^{\ri Z(v+\ri\xi)})     +   \frac{1}{2}\ln\frac{\operatorname{det}(1-\hat{H}^{+})}{\operatorname{det}(1-\hat{H}^{-})}   \notag\\
    &+\frac{1}{2}\int_{-\infty}^{+\infty}dv G(-v-i\xi)\ln(1+e^{\ri Z(v+\ri\xi)})  -\frac{1}{2}\ln(1+e^{\ri Z(0)})  ,  
\end{align}
where $\hat{H}^{ \pm}$ act as
\begin{align}
    \left(\hat{H}^{ \pm}(g)\right)(x)=\oint_{\mathcal{C}} \frac{\mathrm{d} u}{2 \pi \ri} \frac{\mathfrak{a}(u)}{\mathfrak{a}(u)+1} \frac{\varphi(x-u) \pm \varphi(x+u)}{2} g(u)  .   
\end{align}
To use the numerical solution, one should first rescale the integral variable and the imaginary shift $\xi$ by $\tfrac{\gamma}{\pi}$. The first two terms of \eqref{app:gfunct} can be written as  
\begin{align}
    &- \frac{\gamma}{\pi} \operatorname{Im} \int_{-\infty}^{+\infty} \frac{dv \, dx}{2\pi} (\ln f)'(-\tfrac{\gamma}{\pi}v-x-\ri\tfrac{\gamma}{\pi}\xi)  \,  (\delta-G)(x) \ln(1+e^{-\epsilon(v,\xi)})  \notag\\
    &+\frac{1}{2}\frac{\gamma}{\pi} \int_{-\infty}^{+\infty}dv G(-\tfrac{\gamma}{\pi}v-\ri\tfrac{\gamma}{\pi}\xi)\ln(1+e^{-\epsilon(v,\xi)}) .  
\end{align}
The integral can be done by using the convolution theorem, which is the same as the trick used in solving the NLIE. The third term can be calculated by using the equation \eqref{app:nliesol} and take $u$ to be zero. To calculate the the Fredholm determinants, we descritize the action of operator $\hat{H}^{\pm}$ on functions and represent $1-\hat{H}^{\pm}$ as matrices. Before discretization, one should also do the rescaling for the action of $\hat{H}^{ \pm}$, then we have: 
\begin{align}
    \left(\hat{H}^{ \pm}(g)\right)(\tfrac{\gamma}{\pi}x)=\frac{\gamma}{\pi} \oint_{\mathcal{C}} \frac{\mathrm{d} u}{2 \pi i} \frac{\mathfrak{a}(\tfrac{\gamma}{\pi}u)}{\mathfrak{a}(\tfrac{\gamma}{\pi}u)+1} \frac{\varphi(\tfrac{\gamma}{\pi}(x-u)) \pm \varphi(\tfrac{\gamma}{\pi}(x+u))}{2} g(\tfrac{\gamma}{\pi}u) .
\end{align}
Taking the contour $\mathcal{C}$ to be $\{\mathbb{R}+\ri\xi\}\cup\{\mathbb{R}-\ri\xi\}$, then the action becomes that 
\begin{align}
    \left(\hat{H}^{ \pm}(g)\right)(\tfrac{\gamma}{\pi}(x+\ri\xi))=&-\frac{\gamma}{\pi} \int_{-\infty}^{\infty} \frac{\mathrm{d} u}{2 \pi \ri} \frac{e^{-\epsilon(u,\xi)}}{e^{-\epsilon(u,\xi)}+1} \frac{\varphi(\tfrac{\gamma}{\pi}(x-u)) \pm \varphi(\tfrac{\gamma}{\pi}(x+u+\ri2\xi))}{2} g(\tfrac{\gamma}{\pi}(u+\ri\xi))  \notag\\
    &+\frac{\gamma}{\pi} \int_{-\infty}^{\infty} \frac{\mathrm{d} u}{2 \pi \ri} \frac{1}{e^{-\bar{\epsilon}(u,\xi)}+1} \frac{\varphi(\tfrac{\gamma}{\pi}(x-u+2\ri\xi)) \pm \varphi(\tfrac{\gamma}{\pi}(x+u))}{2} g(\tfrac{\gamma}{\pi}(u-\ri\xi)),  \\
    \left(\hat{H}^{ \pm}(g)\right)(\tfrac{\gamma}{\pi}(x-\ri\xi))=&-\frac{\gamma}{\pi} \int_{-\infty}^{\infty} \frac{\mathrm{d} u}{2 \pi \ri} \frac{e^{-\epsilon(u,\xi)}}{e^{-\epsilon(u,\xi)}+1} \frac{\varphi(\tfrac{\gamma}{\pi}(x-u-2\ri\xi)) \pm \varphi(\tfrac{\gamma}{\pi}(x+u))}{2} g(\tfrac{\gamma}{\pi}(u+\ri\xi))  \notag\\
    &+\frac{\gamma}{\pi} \int_{-\infty}^{\infty} \frac{\mathrm{d} u}{2 \pi \ri} \frac{1}{e^{-\bar{\epsilon}(u,\xi)}+1} \frac{\varphi(\tfrac{\gamma}{\pi}(x-u)) \pm \varphi(\tfrac{\gamma}{\pi}(x+u-2\ri\xi))}{2} g(\tfrac{\gamma}{\pi}(u-\ri\xi))
    .
\end{align}
We then discretize these integrals. In doing this, one should be careful that the order of the elements in the matrix should be consistent with the order of points on the integral contour $\mathcal{C}$, where the order of the points on the line $\mathbb{R}+\ri\xi$ should be the inverse of the usual order.

\end{appendix}